\documentclass{amsart}
\usepackage{amssymb,latexsym}
\usepackage{epsfig}
\usepackage{eufrak}
\usepackage{amsmath}
\usepackage{mathrsfs}

\theoremstyle{plain}
\newtheorem{theorem}{Theorem}

\newtheorem{lemma}{Lemma}

\theoremstyle{definition}

\theoremstyle{remark}

\numberwithin{equation}{section}

\newtheorem{remark}{Remark}

\newcommand{\be}{\begin{equation}}
\newcommand{\ee}{\end{equation}}
\newcommand{\bee}{\begin{eqnarray*}}
\newcommand{\eee}{\end{eqnarray*}}
\newcommand{\bel}{\begin{eqnarray}}
\newcommand{\eel}{\end{eqnarray}}
\newcommand{\bec}{\begin{cases}}
\newcommand{\eec}{\end{cases}}
\newcommand{\bem}{\begin{bmatrix}}
\newcommand{\eem}{\end{bmatrix}}

\newcommand{\la}{\label}
\newcommand{\li}{\left}
\newcommand{\ri}{\right}

\newcommand{\DEF}{\stackrel{\mathrm{def}}{=}}

\newcommand{\lf}{\lfloor}
\newcommand{\rf}{\rfloor}

\newcommand{\ep}{\epsilon}
\newcommand{\vep}{\varepsilon}
\newcommand{\lm}{\lambda}

\newcommand{\si}{\sigma}

\newcommand{\de}{\delta}

\newcommand{\vDe}{\varDelta}

\newcommand{\ga}{\gamma}

\newcommand{\ze}{\zeta}

\newcommand{\ba}{\beta}

\newcommand{\ro}{\rho}
\newcommand{\ka}{\kappa}

\newcommand{\f}{\frac}

\newcommand{\cd}{\cdots}
\newcommand{\qu}{\quad}
\newcommand{\qqu}{\qquad}
\newcommand{\fa}{\forall}

\newcommand{\mscr}{\mathscr}
\newcommand{\mcal}{\mathcal}

\newcommand{\bb}{\mathbb}

\newcommand{\wh}{\widehat}

\newcommand{\mrm}{\mathrm}
\newcommand{\bs}{\boldsymbol}

\newcommand{\ap}{\approx}

\newcommand{\sh}{\slash}

\newcommand{\tx}{\text}

\newcommand{\iy}{\infty}

\newcommand{\leu}{\subseteq}

\newcommand{\im}{\imath}
\newcommand{\jm}{\jmath}
\newcommand{\pa}{\partial}

\newcommand{\bed}{\begin{description}}
\newcommand{\eed}{\end{description}}
\newcommand{\bei}{\begin{itemize}}
\newcommand{\eei}{\end{itemize}}
\newcommand{\ben}{\begin{enumerate}}
\newcommand{\een}{\end{enumerate}}
\newcommand{\bib}{\bibitem}
\newcommand{\beL}{\begin{lemma}}
\newcommand{\eeL}{\end{lemma}}
\newcommand{\beT}{\begin{theorem}}
\newcommand{\eeT}{\end{theorem}}
\newcommand{\sect}{\section}

\newcommand{\bpf}{\begin{pf}}
\newcommand{\epf}{\end{pf}}
\newcommand{\bsk}{\bigskip}

\newcommand{\pfbox}{\hfill\mbox{$\Box$}}
\newenvironment{pf}{\paragraph*{Proof{\rm.}}}{\pfbox\bigskip}

\begin{document}

\title[Probabilistic Robustness Analysis]
{Probabilistic Robustness Analysis --- Risks, Complexity and
Algorithms}

\author{Xinjia Chen, Kemin Zhou and Jorge L. Aravena}

\address{Department of Electrical and Computer Engineering\\
Louisiana State University\\
Baton Rouge, LA 70803}

\email{chan@ece.lsu.edu\\
kemin@ece.lsu.edu\\
aravena@ece.lsu.edu}

\thanks{This
research was supported in part by grants from NASA (NCC5-573), LEQSF (NASA /LEQSF(2001-04)-01), the NNSFC Young
Investigator Award for Overseas Collaborative Research (60328304) and a NNSFC grant (10377004).}


\date{August 2006}

\begin{abstract}

It is becoming increasingly apparent that probabilistic approaches
can overcome conservatism and computational complexity of the
classical worst-case deterministic framework and may lead to designs
that are actually safer. In this paper we argue that a comprehensive
probabilistic robustness analysis requires a detailed evaluation of
the robustness function and we show that such evaluation can be
performed with essentially any desired accuracy and confidence using
algorithms with complexity linear in the dimension of the
uncertainty space. Moreover, we show that the average memory
requirements of such algorithms are absolutely bounded and well
within the capabilities of today's computers.

In addition to efficiency, our approach permits control over statistical sampling error and the error due to
discretization of the uncertainty radius. For a specific level of tolerance of the discretization error, our
techniques provide an efficiency improvement upon conventional methods which is inversely proportional to the
accuracy level; i.e., our algorithms get better as the demands for accuracy increase.

\end{abstract}

\maketitle

\sect{Introduction}

In recent years, a number of researchers have proposed probabilistic control methods for overcoming the
computational complexity and conservatism of the deterministic worst-case robust control framework (e.g.,
\cite{bai}--\cite{C3}, \cite{PT}--\cite{Wang} and the references therein).

The philosophy of probabilistic control theory is to sacrifice cases
of extreme uncertainty. Such paradigm has lead to the concept of
{\it confidence degradation function} (originated by Barmish, Lagoa
and Tempo \cite{BLT}), which has demonstrated to be extremely
powerful for the robustness analysis of uncertain systems. Such
function, $\mscr{P}(.)$, is defined as $\mscr{P}(r) = \inf_{0 < \ro
\leq r} \bb{P}(\ro)$ with
\[
\bb{P}(\rho) = \mrm{vol}\{\tx{$X \in \mcal{B}_\rho \mid $ The robustness requirement is guaranteed for $X$}\}
\slash \mrm{vol} \{ \mcal{B}_\rho \} \] where the volume function $\mrm{vol} \{.\}$ is the Lebesgue measure, and
$\mcal{B}_\rho$ denotes the uncertainty bounding set with radius $\rho$.  Interestingly, it was discovered in
\cite{BLT} that such function is not necessarily monotone decreasing in the uncertainty radius. In view of this
fact and for the purpose of avoiding the confusion with the concept of {\it confidence band}, used in the
evaluation of the accuracy of the estimate of $\bb{P}(r)$, the confidence degradation function is referred to as
the {\it robustness function} in this paper. Accordingly, a graph representation of the robustness function is
called the {\it robustness curve}. It can be seen that the robustness function is a natural extension of the
concept of robustness margin. From the robustness curve, one can determine the {\it probabilistic robustness
margin} \cite{BLT} and estimate the deterministic robustness margin.

In addition to overcoming the NP hard complexity and conservatism of
deterministic robustness analysis methods, the robustness function
can address very complex problems which are intractable by
deterministic worst-case methods. Moreover, the probability that the
robustness requirement is guaranteed can be inferred from the
robustness function, while the deterministic margin losses the
connection with such probability. Based on the assumption that the
density function of uncertainty is radially symmetric and
non-increasing with respect to the norm of uncertainty, it has been
shown in \cite{BLT} that the probability that the robustness
requirement is guaranteed is no less than $\mscr{P}(r) = \inf _{\rho
\in (0, r]} \mathbb{P}(\rho)$ when the uncertainty is included in a
bounding set with radius $r$. The underlying assumption is supported
by modeling and manufacturing considerations that the uncertainty is
unstructured so that all directions are equally likely and that
small perturbations from the nominal model are more likely than
large perturbations. Since $\bb{P}(.)$ is not monotonically
decreasing \cite{BLT}, the lower bound of the probability depends on
$\mathbb{P}(\rho)$ for all $\rho \in (0, r]$. It is not clear
whether it is feasible to estimate $\mscr{P}(r)$ since the
estimation of $\mathbb{P} (\rho)$ for every $\rho$ relies on
intensive Monte Carlo simulation and $\bb{P}(\ro)$ needs to be
estimated for numerous values of $\ro$. For such probabilistic
method to overcome the NP hard of worst-case methods, it is
necessary to show that the complexity for estimating $\mscr{P}(r)$
for a given $r$ is polynomial in terms of computer running {\it
time} and memory {\it space}. In this paper, we demonstrate that the
complexity in terms of space and time is surprisingly low and is
{\it linear in the uncertainty dimension and the logarithm of the
relative width of the range of uncertainty radius}.

In the next section we argue that both the deterministic robustness
margin and its risk-adjusted version -- the probabilistic robustness
margin have inherent limitations. We address those limitations
through the use of the robustness function that can describe the
performance of a system over a wide range of uncertainties. In order
to construct the robustness function for wide range of uncertainty
radii, the conventional method independently estimate $\bb{P}(r_i)$
for each grid points of uncertainty. If there are $m$ grid points
and $N$ is the sample size for each radius, then the total number of
simulations is $N m$. In Section \ref{boundS}, we use the sample
reuse principle and demonstrate that the robustness curve for
arbitrarily wide range of uncertainty radii can be accurately
constructed with surprisingly low complexity. Clearly, the number of
grid points, $m$, must tend to infinity as the tolerance tends to
zero. However, we show that with our algorithms, the {\it equivalent
number of grid points} (ENGP), $m_{eq}$, is strictly bounded from
above in the sense that in order to guarantee the same level of
accuracy for the estimation of the robustness function,  the
required average computational effort is  the same as that of a
conventional grid with $m_{eq}$ points.  Moreover, we show that the
average memory requirement  is also absolutely bounded and is well
within the reach of modern computers.

The remainder of the paper is organized as follows. Section
\ref{riskS} provides an example illustrating the pitfalls of
deterministic robustness margin and the probabilistic robustness
margin. Section \ref{errorS} discusses the control of estimation
error of the robustness function and the required complexity.
Section \ref{Difficulties} investigates the difficulties of the
conventional data structure. Section \ref{algorSM} describes our
new algorithms, analyzes the complexity of data processing and
memory space, and introduces the concept of confidence band. The
proofs of all the theorems are included in the Appendices.

\section{The Risk of Robustness Margins} \label{riskS}

In this section we make the case for the need to have a robustness
function in order to properly estimate how well a control system
tolerates uncertainties. Conventional robust control approaches the
issue with a ``worst case'' philosophy. In this regard, it has been
demonstrated (Chen, Aravena and Zhou, \cite{CJZ}) that it is not
uncommon for a probabilistic controller to be significantly less
risky than a deterministic worst-case control. The reasons are the
``uncertainty in modeling uncertainties'' and the fact that the
worst-case design cannot, in some instances, be ``all
encompassing.'' Therefore, the worst-case approach has an associated
risk that usually is overlooked, while the probabilistic approach
acknowledges the risk and manages it.

From manufacturing and modeling considerations, it is sensible to
assume that the density of the distribution of uncertainty decreases
with increasing uncertainty norm. Such assumption leads to the
worst-case property of uniform distribution in robustness analysis
\cite{BLT}. However, the decay rate of density is generally unknown
to the designer. Therefore, for a given uncertainty radius $r$, one
does not have good knowledge about the coverage probability of the
uncertainty set $\mcal{B}_r$. It is important to note that the
system robustness depends critically on the distribution of
uncertainty norm.

\begin{figure}[htb]
\centering{
\setlength{\unitlength}{0.0006in}%
\begin{picture}(4224,1074)(1189,-3523)
\thicklines \put(1801,-2761){\circle{150}}
\put(2401,-3061){\framebox(600,600){$C$}} \put(1201,-2761){\vector(
1, 0){525}} \put(1876,-2761){\vector( 1, 0){525}} \put(3001,-2761){\vector(
1, 0){900}}
\put(3901,-3061){\framebox(600,600){$G$}} \put(4501,-2761){\vector( 1,
0){900}} \put(4801,-2761){\line(
0,-1){750}} \put(4801,-3511){\line(-1, 0){3000}} \put(1801,-3511){\vector(
0, 1){675}}
\put(1276,-2611){\makebox(0,0)[lb]{$r$}}
\put(2026,-2611){\makebox(0,0)[lb]{$e$}}
\put(5101,-2611){\makebox(0,0)[lb]{$y$}}
\put(3451,-2611){\makebox(0,0)[lb]{$u$}}
\put(1551,-3136){\makebox(0,0)[lb]{$-$}}
\end{picture}
\caption{Standard Feedback Configuration} \label{fig108} }
\end{figure}
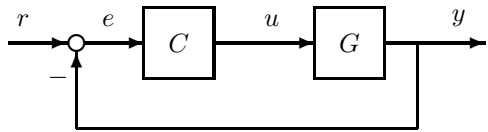

Attempts to improve the analysis have led to the definitions of a
{\em deterministic robustness margin} and a {\em probabilistic
robustness margin}. Both are numbers that purportedly allow the user
to estimate the tolerance to uncertainties. We contend that both can
be misleading, and for essentially the same reason. To demonstrate
this view point, we consider a feedback system shown in Figure
\ref{fig108}.

The transfer function of the plant is $G(s) = \f{q}{s - p}$ where
$p$ and $q$ are uncertain parameters. The uncertainty bounding set
with radius $r > 0$ is
\[
\mcal{B}_r = \{ (x, y): \; |x - q_0| \leq r, \qu |y - p_0| \leq r
\}, \qqu p_0 < 0, \qqu q_0 > 0.
\]
Consider two controllers $C_A = \f{ K_A } { s + \si }, \qu \si
>0$ and $C_B = K_B$ such that
\[
\la{con_b} 1 < K_B < \f{K_A}{\si}, \qqu \f{ K_A \; q_0 - \si \; p_0
} {K_A + \si} < \si - p_0. \] Suppose that the robustness
requirement is stability. It can be shown that the robustness
function for controller $A$ is
\[
\mscr{P}^A (r) = \bec 1 & \tx{for} \qu 0 < r < \ro_A;\\
\f{1}{2} - \f{ K_A \li ( r + \f{\si \; \ba}{K_A} - q_0 \ri )^2 } { 8
\si \; r^2} - \f{ p_0 - \ba } {2 r } & \tx{for}
\qu \ro_A \leq r \leq \ro_A^*;\\
\f{1}{2} - \f{ (r + \ba - p_0) \li ( r + \f{\si ( \ba + p_0 - r)} {2
K_A} - q_0 \ri ) } { 4 r^2 } - \f{ (p_0 - \ba) } { 2 r } & \tx{for}
\qu r > \ro_A^* \eec \] where \[ \ro_A = \f{ K_A \; q_0 - \si \; p_0
} {K_A + \si}
\]
is the deterministic robustness margin, $\ba = \min (\si, p_0 + r)$,
and $\ro_A^* = \f{ K_A \; q_0 - \si \; p_0 } {K_A - \si}$.  It can
be shown that the robustness function for controller $B$ is given by
\[ \la{pro}
\mscr{P}^B (r) = \bec 1 & \tx{for} \qu 0 < r < \ro_B;\\
1 - \f{ K_B \li ( r + \f{p_0 + r}{K_B} - q_0 \ri )^2 } { 8 r^2} &
\tx{for}
\qu \ro_B \leq r \leq \ro_B^*;\\
\f{1}{2} - \f{\f{p_0}{K_B} - q_0} { 2 r } & \tx{for} \qu r > \ro_B^*
\eec \] where
\[
\ro_B = \f{ K_B \; q_0 - p_0 } {K_B + 1}
\]
is the deterministic robustness margin and $\ro_B^* = \f{ K_B \; q_0 - p_0 }
{K_B - 1}$.

\bsk

\begin{figure}[htbp]
\centerline{\psfig{figure=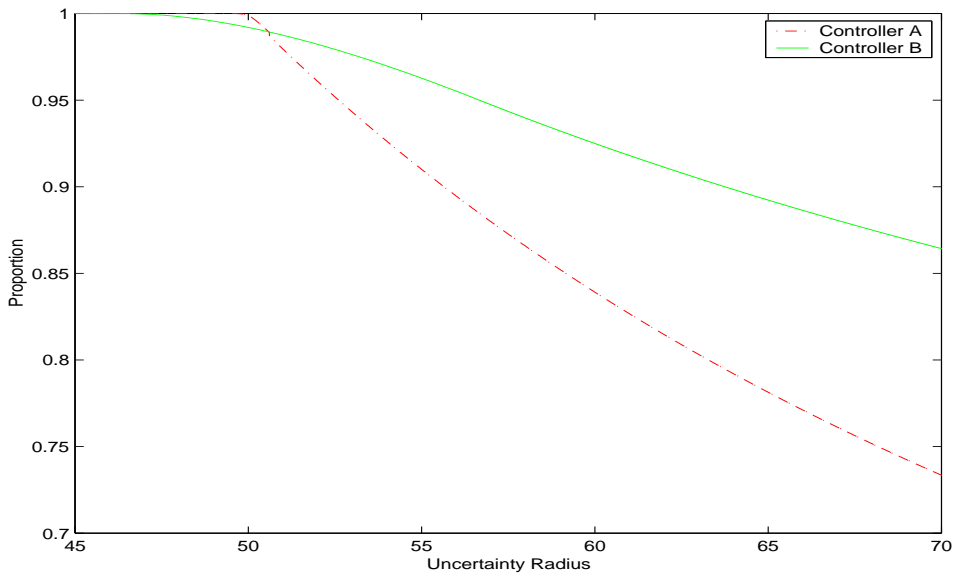, height=3.0in,width=5.0in}}
\caption{Comparison of Controller Alternatives.
\label{fig01}}
\end{figure}

We consider an example with $p_0 = -10, \; q_0 = 50, \; \si = 40, \; K_A = 100 \si, \; K_B = 10$. The
corresponding robustness functions are displayed in Figure \ref{fig01}. We obtained deterministic margins $\ro_A
= 49.6040, \; \ro_B = 46.3636$. Since $\ro_A > \ro_B$, a comparison based on the deterministic margin simply
suggests that controller $A$ is more robust than controller $B$. Quite contrary, a judgement based on the
robustness curves indicates that controller $B$ may be more robust. The risk of the probabilistic robustness
margin can also be illustrated by this example.

Robust analysis should be able to help a designer to reliably
determine which controller design is more robust. However, it
appears that the concepts of robustness margin fail to meet such
fundamental needs of control engineering. On the other hand, the
robustness curve serves the purpose of giving the designer
complete information on how well a control system tolerates
uncertainties.

From the previous discussion, it can be seen that
there are two crucial factors to be considered in order to make a
reliable judgment about the system robustness:

\bed

\item [(i)] How fast the robustness curve rolls off.

\item [(ii)] The dependency of coverage probability of uncertainty
bounding set $\mcal{B}_{r}$ on the radius $r$.

\eed

The second factor can be a difficulty since a designer generally lacks knowledge of the coverage probability
corresponding to a bounding set of fixed radius. To overcome such difficulty, the only choice is to construct
the robustness curve for a wide range of uncertainty radius. The construction of the robustness curve may be
seen as a computationally challenging task since the probability of guaranteeing robustness requirement needs to
be estimated for many values of uncertainty radius. However, as we demonstrate in the next section, using the
sample reuse principle one can construct the robustness curve for virtually the entire scope of uncertainty
range $(0, \iy)$ with absolutely bounded average computational requirements, regardless of the size of the grid.
For example, we shall show that for an uncertainty range as large as $(10^{-10}, \; 10^{10})$, in the average,
one needs less than 50 times memory and computational resources than those needed to evaluate the uncertainty
range $(1, \; e)$ with the same resolution.

\sect{Equivalent Number of Grid Points}\label{boundS}

Throughout this paper, we assume that the uncertainty sets are
homogeneous star-shaped (e.g., \cite{BLT}). That is, the uncertainty
bounding set with radius $r$ is $\mathcal{B}_r = \{ r X \mid X \in
\mathcal{B}_1 \}$ where $\mathcal{B}_1$ denotes the uncertainty
bounding set such that $c X \in \mathcal{B}_1$ for any $X \in
\mathcal{B}_1$ and any $c \in [0,1]$. Clearly, most of the commonly
used uncertainty bounding sets such as the $l_p$ balls and spectral
norm balls are homogeneous star-shaped.

We shall consider the problem of constructing the robustness curve
for arbitrary robustness requirement $\mathbf{P}$ under such
assumption of uncertainty sets. Conventionally, the robustness curve
for a range of uncertainty radii $\li [ \f{a} {\lm}, \; a \ri ]$
with $a > 0, \; \lm > 1$ is constructed by choosing a set of grid
points $\f{a}{\lm} = r_1 < r_2 < \cd < r_m = a$ and, for every grid
point, performing $N$ i.i.d. Monte Carlo simulations. Hence, the
total number of simulations is a deterministic constant $m N$. To
reduce computational complexity, we shall make use of the following
intuitive concept:

\bsk

{\it Let $X$ be an observation of a random variable with uniform
distribution over $\mcal{ B}_\ro \supseteq \mcal{ B}_r$ such that $X
\in \mcal{ B}_r$. Then $X$ can also be viewed as an observation of a
random variable with uniform distribution over $\mcal{B}_r$. }

\bsk

In order to apply such concept, it is necessary to perform the
simulation in a backward direction so that appropriate evaluations
of the robust requirement for larger uncertainty sets can be saved
for the use of later simulations on smaller uncertainty sets
\cite{C0}. The sample reuse principle allows a single simulation to
be used for multiple radii. Thus, the actual total number of
simulations is significantly reduced. In order to quantify this
reduction we introduce the {\em equivalent number of grid points}
(ENGP), $m_{eq}$, defined as \[ m_{eq}=\frac{\mbox{expected total
number of simulations}}{N}. \]

In our approach, the number of simulations required at uncertainty
radius $r_i$, denoted by ${\bf n}_i$ for $i = 1, \cd, m$, is a
random number. The total number of simulations can be represented by
the random variable $\bs{n} = \sum_{i=1}^m {\bf n}_i$. The expected
value of the total number of simulations is $\bb{E} [ \bs{n} ] =
\sum_{i=1}^m \bb{E} [ {\bf n}_i]$ where $\bb{E}[X]$ denotes the
expectation of random variable $X$. Hence, we can formally define \[
m_{eq}=\frac{\bb{E} [ \bs{n} ]}{N}. \] Due to sample reuse, we can
achieve a substantial reduction of simulations, i.e., $\bb{E} [
\bs{n} ] << m N$. To quantify the reduction of the computational
effort, we have introduced the notion of {\it sample reuse factor}
\cite{C0}, which is defined as \be \la{eqreuse} \mcal{
F}_\mathrm{reuse} \DEF \frac{ m N } { \bb{E} [ \bs{n} ]
}=\frac{m}{m_{eq}}. \ee

In our approach, $N$ i.i.d simulation results are collected for each grid
point. Hence, the accuracy of
estimation is the same as that of the conventional method. However, the
average number of simulations in our
approach is $\bb{E} [ \bs{n} ]$, which is equivalent to the complexity of
$m_{eq}$
grid points in the conventional scheme.
As a direct consequence of Theorem 1 of \cite{C0}, we have that, for
any discretization scheme, $m_{eq}$ is independent of the sample
size $N$. Moreover, we have the following general results.

\beT
\la{Bound_General} Let $d$ be the dimension of uncertainty parameter space.
Then, for arbitrary gridding scheme,
the equivalent number of grid points based on the principle of sample reuse
is strictly bounded from above by $1
+ d \; \ln \lm$, i.e.,
\[
m_{\mrm{eq}} < 1 + d \; \ln \lm .
\]
\eeT

See Appendix A for a proof. By an ``arbitrary'' discretization
scheme, we mean two things: i) the number of grid points can be
arbitrarily large; ii) the grid points can be distributed
arbitrarily over the specified range of uncertainty radius.

A fundamental question of robust control is whether randomized
algorithms have polynomial complexity. In light of the fact the cost
of each simulation depends on problem cases, the computational
complexity is usually measured in terms of the number of
simulations. This theorem reveals the following important facts:

\begin{description}

\item [(a)] The complexity is linear in the dimension of the
uncertainty space. Thus our algorithms overcome the curse of
dimensionality.

\item [(b)] The complexity depends linearly in the logarithm of
the ``relative'' width, $\lambda$, of the interval of uncertainty
radii. This proves that our algorithms are capable of estimating
the robustness function for a wide range of uncertainty.

\item [(c)] Our algorithms can arbitrarily reduce the grid error,
while keeping the complexity strictly below a constant bound.

\end{description}

In order to illustrate these points, Figure \ref{fig_2} displays the
variation of $m_{eq}$ for various dimensions of the uncertainty
space and for values of $\lambda$ up to $\lm = 10^{20}$
corresponding to the uncertainty range $(10^{-10}, 10^{10})$ (which
may be deemed a good approximation to $(0, \iy)$). Notice that even
for dimensions as high as $d=1024$ the equivalent number of grid
points, $m_{eq}$, is very reasonable.

\begin{figure}[htbp]
\centerline{\psfig{figure=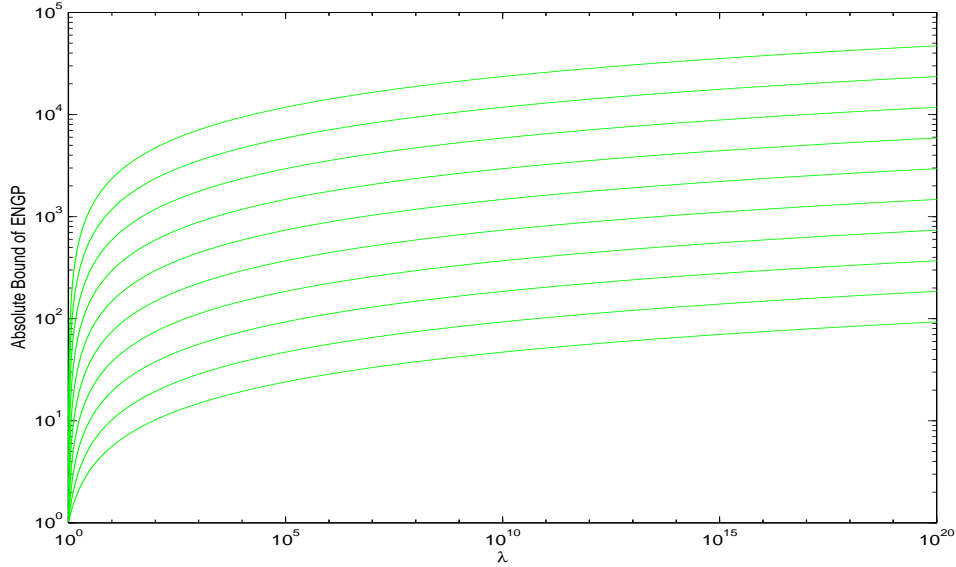, height=3.0in, width=5.0in }}
\caption{Absolute Bounds for $m_{eq}$ (ENGP) ($d = 2^i, \; i = 1, \cd, 10$). }
\la{fig_2}
\end{figure}

Finally in this section, we consider the case where we need to
estimate $\bb{P}(r)$ for $r \in [\ga, U]$ where $\ga > 0$ is a
constant, and $U$ is an estimate of the probabilistic robustness
margin calculated by randomized algorithms. Clearly, $U$ is a random
variable. If $U$ depends on samples which are independent of the
samples generated from the uncertainty set with radius $r \in [\ga,
U]$ we have the following result:

\bsk For any gridding scheme, \be \la{genn} m_{eq} < 1 + d \ln
\f{\bb{E} [U] } { \ga }. \ee

\bsk

To prove (\ref{genn}), notice that $\bb{E} [ \bb{E} [ X \mid Y] ] =
\bb{E} [X]$ for any random variables $X$ and $Y$. Hence, by Theorem
\ref{Bound_General}, \bee m_{\mrm{eq}} & = & \bb{E} [ \bb{E} [
m_{\mrm{eq}} \mid U ] ] < \bb{E} \li [ 1 + d \ln \f{U} { \ga } \ri ]
= 1 + d \; \bb{E} \li [ \ln \f{U} { \ga } \ri ] < 1 + d \ln
\f{\bb{E} [U] } { \ga } \eee where the last inequality is obtained
from applying the Jensen's inequality to the concave function
$\ln(.)$.

\bsk

\sect{Error Control} \label{errorS}

In addition to efficiency, another important issue in any
numerical approach is error control. This point has been
emphasized in many control engineering problems. For instance,
when computing the $H_\iy$ norm of a system, a lower bound and an
upper bound are obtained and is required that the gap between them
be less than a prescribed tolerance. A similar situation arises in
the computation of the structured singular value $(\mu)$.

For the specific case of the estimation of the robustness function, there are two sources of error: i) the
statistical sampling error due to the finiteness of the sample size, $N$ (sample size error); ii) the
discretization error due to the finite number of points in any partition. Control of the sample size error has
been well studied and emphasized. Existing techniques include the Chernoff bounds \cite{Chernoff}, binomial
confidence interval \cite{C3, Clo}, etc. However, we claim that control of discretization error is not
sufficiently emphasized. In fact, one can argue that {\it controlling the sample size error can be meaningless
if the discretization error is not controlled}. This will be the case, for example, for those situations where a
risk at the level of a small $\vep$ (e.g., $\vep = 0.001$) may be significant or unacceptable. {\it How can any
estimation be useful if the discretization error is not ensured to be less than the tolerance $\vep$ }?

In this section, we first introduce an interpolation result
necessary to analyze error control methods. Afterward, we discuss
two different schemes which insure a discretization error less than
a given $\ep \in (0,1)$. The first is a uniform partition whereby
the uncertainty radius interval $[ \f{a}{\lm}, a]$ is partitioned by
$m$ points \be \la{def2} r_i = a - \f{ (m - i) (\lm - 1) }{ (m -1)
\lm } a, \qqu i = 1, \cd, m. \ee In the second scheme we consider a
geometric type partition of the form

\be \la{def} r_i = a \li ( \f{1}{\lm} \ri )^ {\f{m - i} {m-1} },
\qqu i = 1, \cd, m. \ee

For any partition of the uncertainty radius interval, we have the
following linear interpolation results.

\beT \la{bound1} Given an arbitrary partition of the uncertainty
radius interval $\li[ \f{a}{\lm}, \; a \ri ]$ with $\f{a}{\lm} = r_1
< r_2 < \cd < r_m = a$, define
\[
\bb{P}^* (r) = \f{ (r - r_i) \; \bb{P} (r_{i+1}) + (r_{i+1} - r ) \;
\bb{P} (r_{i}) } { r_{i+1} - r_i }, \]
\[
g(r) = (r_{i+1} - r) \li ( \f{r}{r_i} \ri )^{-d} + (r - r_{i}) \li (
\f{r_{i+1}}{r} \ri )^{-d}. \]  Then, for all $r \in [r_i, r_{i+1}]$,
\[ | \bb{P} (r) - \bb{P}^* (r) | \leq 1 - \f{ g(r_\star) } { r_{i+1} - r_i }
\leq \f{d}{2 r_i} (r_{i+1} - r_i)
\]
where $r_\star \in (r_i, r_{i+1})$ is the unique solution of
equation
\[
\li ( \f{r_{i+1}}{r} \ri )^{-d} \li [ 1 + \li ( 1 - \f{r_i} { r }
\ri ) d \ri ] - \li ( \f{r}{r_{i}} \ri )^{-d} \li [ 1 + \li (
\f{r_{i+1}} { r } - 1 \ri ) d \ri ] = 0
\]
with respect to $r$, which can be solved by a bisection search.
\eeT

See Appendix B for a proof. As mentioned before, these interpolation
results will be used in the construction of a tight confidence band
for the robustness function.

\begin{remark}
To guarantee a prescribed tolerance $\ep \in (0,1)$, the number of
grid points must be larger than a certain number. It has been shown
by Barmish, Lagoa and Tempo \cite{BLT} that if \be \la{Bar_La_Te} m
\geq 1 + \f{ 2 (\lm -1) d } { \ep } \ee then $|\bb{P} ( r ) -
\bb{P}( r_i)| < \ep \qu \fa r \in [r_i, r_{i+1}]$ for $i = 1, \cd,
m-1$. This bound shows that, for fixed error $\ep$, the complexity
is polynomial. From another perspective, it also shows that the
number of grid points and computational complexity tend to infinity
as the tolerance tends to zero. For example, the robustness analysis
problem for complex uncertainty of size $30 \times 30$ over an
interval of uncertainty with $\lm = 10$, requires $m \geq
3,240,000,001$ in order to guarantee $\ep \leq 10^{-5}$. The bound,
however, does not account for the sample reuse principle. Using our
approach the equivalent number of grid points for this case is
bounded from above by $1 + 1800 \times \ln(10)$.

\end{remark}

The following result is our extension of the result by Barmish et al., cited
above, and quantifies the advantage
of using linear interpolation.

\beT \la{grid_uni} Let \be \la{grid_uniform} m = 2 + \li \lf \f{
(\lm -1) d } { 2 \ep } \ri \rf \ee where $\lf . \rf$ denotes the
floor function. Then, for a uniform gridding scheme,
\[
|\bb{P} ( r ) - \bb{P}^* ( r)| < \ep \qqu \fa r \in [r_i, r_{i+1}]
\]
for $i = 1, \cd, m-1$. Moreover, the equivalent number of grid
points is
\[
m_{\mrm{eq}} (\epsilon) = m - \sum_{i=1}^{m-1} \li ( 1 - \f{1} {
\f{m-1} {\lm-1} + i } \ri )^d.
\]
\eeT

See Appendix C for a proof. \begin{remark} We point out that when
using linear interpolation the number of grid points given by
(\ref{grid_uniform}) is approximately $\f{1}{4}$ of the bound given
by (\ref{Bar_La_Te}).
\end{remark}

We now analyze a discretization scheme whereby the partition of the
uncertainty interval under study is defined by a geometric series.

\beT \la{Grid_geometric} For a geometric discretization scheme with
\[
m = 2 + \li \lf \f{ \ln \lm } { \ln \li ( 1 + \f{2 \ep}{d} \ri ) }
\ri \rf
\]
and \[ r_i = a \li ( \f{1}{\lm} \ri )^ {\f{m - i} {m-1} } \] for $i
= 1, \cdots, m$, the following statements hold true:

\bed

\item [(I)]
\[
|\bb{P} ( r ) - \bb{P}^* ( r)| < \ep \qqu \fa r \in [r_i, r_{i+1}],
\qqu i = 1, \cd, m - 1.
\]

\item [(II)]
\[
m_{\mrm{eq}}(\epsilon) = 1 + \li ( 1 + \li \lf \f{ \ln \lm } { \ln
\li ( 1 + \f{2 \ep}{d} \ri ) } \ri \rf \ri ) \; \li [ 1 - \left(
\frac{1}{\lm} \right)^{ \f{d}{ 1 + \li \lf \f{ \ln \lm } { \ln \li (
1 + \f{2 \ep}{d} \ri ) } \ri \rf} } \ri ].
\]

\item [(III)]
\[
\mcal{ F}_\mathrm{reuse} > \f{1}{2 \ep} \li ( 1 - \f{1}{ 1 + d \;
\ln \lm } \ri ).
\]

\eed

\eeT

See Appendix D for a proof.

\begin{remark}
Since $ 1 + d \; \ln \lm >> 1$ in many situations, the sample reuse
factor for the geometric discretization scheme may be written in a
more elegant form. That is,
\[
\mcal{ F}_\mathrm{reuse} \ap \f{1}{2 \ep}
\]
which is inversely proportional to the tolerance of the
discretization error. For example, to ensure that the discretization
error is less than $10^{-4}$, which is a rather weak requirement for
many applications, our algorithm reduces the computational effort by
a factor of $5,000$ when compared to a conventional approach.

The two discretization schemes considered here, and others, have bounded
complexity, but the distributions of
the total number of simulations are different. Hence it is reasonable to ask
if there is a ``best
discretization.'' Our results indicate that the geometric scheme is
generally more efficient, as shown by the
comparison of grid points in Figure \ref{fig012} and the comparison of ENGP
in Figure \ref{fig011}.

\begin{figure}[htbp]
\centerline{\psfig{figure=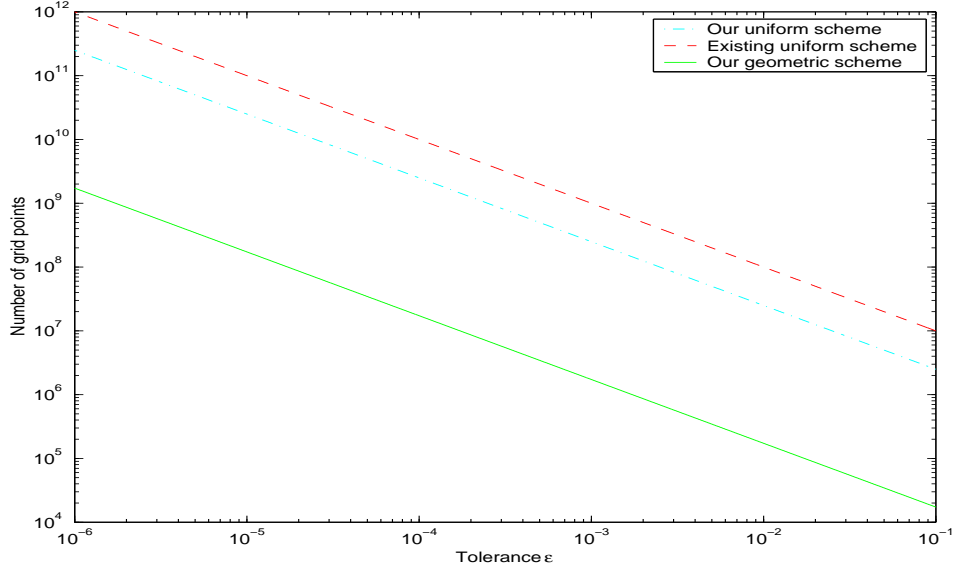, height=3.0in,
width=5.0in }} \caption{Comparison of Number of Grid Points ($\lm
= 10^3, \; d = 500$) } \la{fig012}
\end{figure}

\begin{figure}[htbp]
\centerline{\psfig{figure=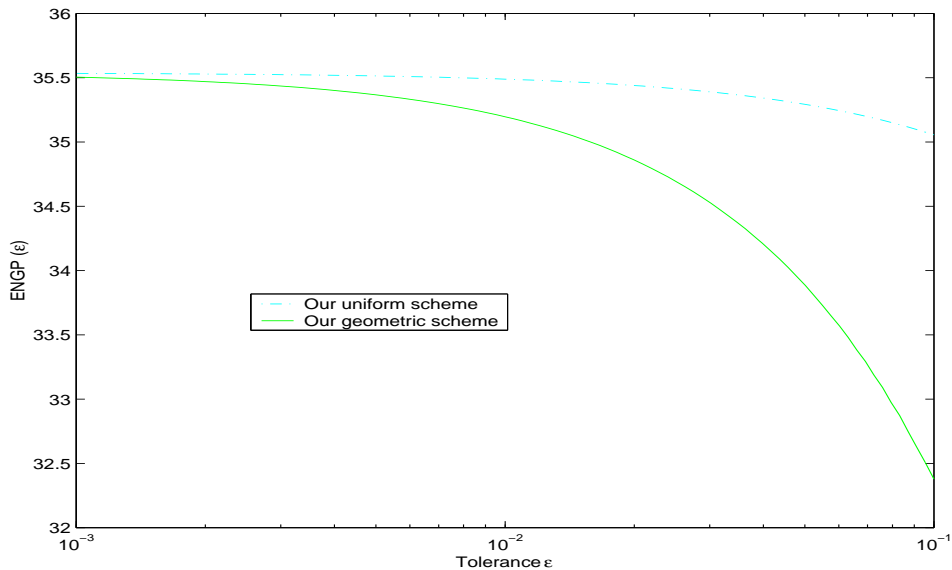, height=3.0in,
width=5.0in }} \caption{Comparison of $m_{eq}$ ($\lm = 10^3, \; d
= 5, \; 1 + d \ln \lm = 35.5388$) } \la{fig011}
\end{figure}

\end{remark}

\bsk

\sect{The Difficulties of Conventional Data Structure}
\label{Difficulties}

Our previous sample reuse algorithm \cite{C0} uses the same data
structure as that of the conventional algorithm. That is, the data
structure for implementing the algorithms is basically a matrix of
fixed size. In such data structure, for each grid point $r_i$, there
is a record $(k_i, n_i)$ where $k_i$ represents the number of cases
guaranteeing (or violating) the robustness requirement among $n_i$
simulations. In the course of experiment, the number $n_i$ is
increment from $0$ to sample size $N$. In the following two
subsections, we demonstrate that the conventional data structure is
not suitable for controlling the error due to finite gridding.

\subsection{The Issue of Data Processing}

Clearly, the total number of records is exactly the number of grid points $m$. For the conventional method, to
accomplish $N$ simulations for each grid points, the total number of updating the data record is $N m$. As
illustrated in Section 4, to control the error due to finite gridding requires an extremely large number, $m$,
of grid points even for moderate requirement of $\ep$. Therefore, $Nm$ is usually a {\it very large number}. It
can be shown that {\it if the sample reuse algorithm employs the same data structure as that of the conventional
method, then, for any gridding scheme with $m$ grid points, the total number of times of updating the data
record is also $N m$.} This is true because, for every time a record $(k_i, n_i)$ is updated, the number $n_i$
can only be increased by $1$, and the number $n_i$ must be $N$ when the experiment is completed. To have a
feeling that the data processing with the conventional data structure is a severe challenge, one can consider
the example discussed in Remark 1 of Section 4. With $m \geq 3, 240, 000, 001$ and normal sample size $10^4 < N
< 10^6$, it can be seen that $Nm$ will be in the range of $3 \times 10^{13}$ to $3 \times 10^{15}$. This is an
enormous burden for today's computing technology. For a modern computer with $1.9$ GHz CPU and $256$ M bytes
RAM, it takes about $20$ seconds to execute $10^7$ times the command $n_i \leftarrow n_i + 1$ written in the
MATLAB language. It can be reasonably inferred that updating the data record for $3 \times 10^{13}$ times will
take about $20 \times 10^{-7} \times 3 \times 10^{13}$ seconds (i.e., about $700$ days).

\subsection{The Issue of Memory Space}

For the conventional data structure, the total number of records is $m$. To execute the sample reuse algorithm
or the conventional one with such data structure, each record must occupy some physical addresses. Such
addresses are necessary for storing and visualizing the outcome of simulations. Of course, to obtain the outcome
simulations may require a much higher amount of computer internal memory to execute the algorithm. Since $m$ is
usually a very large number, the consumption of memory to store and visualize the output of simulation can be
enormous. To illustrate, consider again the example discussed in Remark 1 of Section 4. Since a floating point
number occupies $2$ bytes, storing a tuple of the form $(k_i, n_i)$ needs $4$ bytes. For $m \geq 3, 240, 000,
001$, the data record will consumes $4 \times 3, 240, 000, 001 \ap 13 \times 10^9$ bytes (i.e., about $13$ giga
bytes) of RAM. Such requirement, just for visualizing the outcome of the simulations, is a challenging task even
for modern computers.

\sect{New Techniques of Sample Reuse} \label{algorSM}

In the last section, we have shown that any algorithm using the
conventional data structure suffer from the problems of the
complexity of data processing and memory space. This is because, the
sample size $N$ is usually very large and the number, $m$, of points
in the partition of uncertainty radius approaches infinity as the
tolerance, $\ep$, approaches zero (see Theorem \ref{grid_uni}). In
this section, we shall demonstrate that, by introducing a dynamic
data structure and a new sample reuse algorithm, the average
requirement of memory and the computational effort devoted to data
processing are absolutely bounded, independent of the tolerance, and
well within the power of modern computers.

\subsection{Data Structure} In order to address the memory
issue and minimize the effort devoted to data processing, an appropriate data structure is critical. The key
idea is to make use of the observation that, {\it for a set of consecutive grid points with identical records of
simulation results, it suffices to store the information of the smallest and the largest grid points.} To
illustrate our techniques, we enumerate, in a chronicle order of generation, the samples generated from various
uncertainty bounding sets as $X_1, X_2, \cdots$. When samples $X_1, X_2, \cd, X_j$ have been generated, the
state of the experiment is completely represented by functions $\bs{s} (i, j)$ and $\bs{v} (i, j)$, where
\[
\bs{s} (i, j) = \sum_{k=1}^{ \bs{l}(i,j) } Y_i^k, \qqu \bs{v} (i, j)
= \sum_{k=1}^{ \bs{l}(i,j) } Z_i^k
\]
with
\[
Y_i^k = \bec 1 & \tx{if $X_k \in \mathcal{B}_{r_i}$};\\
0 & \tx{otherwise} \eec
\]
\be \la{tru}
\bs{l}(i,j) = \max \li \{\ell : \; 1 \leq \ell \leq j, \; \sum_{k=1}^{\ell}
Y_i^k \leq N \ri \}
\ee and
\[
Z_i^k = \bec 1 & \tx{if $X_k \in \mathcal{B}_{r_i}$ and
$\mathbf{P}$ is violated for $X_k$};\\
0 & \tx{otherwise} \eec
\]
for $i = 1, \cd, m$ and $k = 1, \cd, j$.  The reason we introduce
variable $\bs{l}(i,j)$ by (\ref{tru}) is that, for grid point $r_i$,
once $N$ equivalent simulations are available, the subsequent
simulations can be ignored. By the principle of sample reuse,
$\bs{s} (i, j)$ and $\bs{v} (i, j)$ are, respectively, the
accumulated numbers of samples and violations for uncertainty
bounding set with radius $r_i$. When the experiment is completed, we
have $\bs{n}$ samples $X_1, X_2, \cd, X_{\bs{n}}$ and
\[
\bs{s} (i, \bs{n}) = N, \qqu \bb{P} (r_i) = 1 - \f{ \bs{v} (i,
\bs{n}) } { N}, \qqu i = 1, \cd, m.
\]
It can be seen that $\bs{s} (i, j)$ is piece-wise constant (with
respect to $i$) and there exists a matrix $S^j$ such that, for $i =
1, \cd, m$, \be \la{con1} \bs{s} (i, j) = \bec [S^j]_{\ell, 2} &
\tx{for} \qu [S^j]_{\ell, 1} \leq i
< [S^j]_{\ell + 1, 1} \qu \tx{with} \qu 1 \leq \ell \leq \ka - 1;\\
[S^j]_{\ka, 2} & \tx{for} \qu [S^j]_{\ka, 1} \leq i \leq m \eec \ee
where $\ka$ is the number of rows of $S^j$ and $[A]_{\im,\jm}$
denotes the element of matrix $A$ in the $\im$-th row and the
$\jm$-th column. Roughly speaking, {\it the first column of matrix
$S^j$ records the indexes of grid points for which the accumulated
numbers of samples are jumping to different values. The second
column of matrix $S^j$ records the corresponding accumulated numbers
of samples.}

Similarly, $\bs{v} (i, j)$ is piece-wise constant (with respect to $i$) and
there exists a matrix $V^j$ such
that, for $i = 1, \cd, m$, \be \la{con2}
\bs{v} (i, j) = \bec [V^j]_{\ell, 2} &
\tx{for} \qu [V^j]_{\ell, 1}
\leq i < [V^j]_{\ell + 1, 1} \qu \tx{with} \qu 1 \leq \ell \leq \tau - 1;\\
[V^j]_{\tau, 2} & \tx{for} \qu [V^j]_{\tau, 1} \leq i \leq m \eec
\ee where $\tau$ is the number of rows of $V^j$. Loosely speaking,
{\it the first column of matrix $V^j$ records the indexes of grid
points for which the accumulated numbers of violations are jumping
to different values. The second column of matrix $V^j$ records the
corresponding accumulated numbers of violations.}

In this paper, matrices $S^j$ and $V^j$ are, respectively, referred
to as the {\it matrix of sample sizes} and the {\it matrix of
violations}. At any stage that samples $X_1, \cd, X_j$ have been
generated, the status of the experiment is completely characterized
by matrices $S^j, \; V^j$. Both matrices are of two columns but of
varying number of rows in the course of experiment.

To save memory and data processing effort, we shall take advantage
of the piece-wise constant property of the accumulated numbers of
samples and violations. Hence, we shall construct matrices $S^j$ and
$V^j$ when we have generated $X_1, \cd, X_j$. As can be seen in the
sequel, such matrices can be constructed recursively. Once we have
$S^j$ and $V^j$, we can generate sample $X_{j+1}$ and update $S^j,
\; V^j$ as $S^{j+1}, \; V^{j+1}$ in accordance with equations
(\ref{con1}) and (\ref{con2}).

\subsection{Sample Reuse Algorithm} In this section, we present our sample
reuse
algorithms as follows.

\bed

\item [Initialization] We initialize the matrices of sample sizes
and violations as follows:

$\qu \diamondsuit$ Generate sample $X_1$ uniformly from uncertainty set with
radius $r_m$.

$\qu \diamondsuit$ Compute $\jm$ such that $X_1 \in \mathcal{B}_ {r_i}$ for
$\jm \leq i \leq m$ and $X_1 \notin
\mathcal{B}_ {r_i}$ for $1 \leq i \leq \jm - 1$.

$\qu \diamondsuit$ Let $S^1 = \bem 1 & 1 \eem$ if $\jm = 1$ and $S^1 = \bem
1 & 0\\
\jm & 1 \eem$ if $\jm > 1$.

$\qu \diamondsuit$ Let $V^1 = \bem 1 & 0 \eem$ if $\bb{I} (X_1) = 0$
and $V^1 = S^1$ if $\bb{I} (X_1) = 1$, where $\bb{I}(X) =1$ if the
robustness requirement is violated for $X$ and otherwise $\bb{I}(X)
= 0$.

\item [Sample generation] If $[S^j]_{\ka,1} < N$ then generate
sample $X_{j+1}$ uniformly from uncertainty set with radius $r_m$,
otherwise generate sample $X_{j+1}$ uniformly from uncertainty set
with radius $[S^j]_{\ka - 1,1}$.

\item [Updating matrices] Update $S^j$ as $S^{j+1}$ by the method described
in Section \ref{sec521}.
If $\bb{I}(X_{j+1}) = 0$ then let $V^{j+1} = V^j$,
otherwise update $V^j$ as $V^{j+1}$ by the method described in Section
\ref{sec522}.

\item [Stopping criterion] The sampling process is terminated if $S^j$
has only one row and $[S^j]_{1,2} = N$.

\eed

\subsubsection{Sample Sizes Tracking} \la{sec521}

In this section, we describe how to update the matrix of sample
sizes. The key idea is to ensure condition (\ref{con1}). Let $\ka$
be the number of rows of $S^j$. We proceed as follows.

\bed

\item [Step (1)] Compute an index $\jm^*$ such that $X_{j+1} \in
\mathcal{B}_ {r_i}$ for $\jm^* \leq i \leq m$ and $X_{j+1} \notin
\mathcal{B}_ {r_i}$ for $1 \leq i \leq \jm^* -
1$ (Note that explicit formulas for computing $\jm^*$ are available when
using uniform or geometric grid
scheme).

\bsk

\item [Step (2)] Modify $S^j$ as a temporary matrix $\wh{S}^{j+1}$
based on the following three cases.

\bsk

Case (1):  $[S^{j}]_{\ell^*,1} < \jm^* < [S^{j}]_{\ell^* + 1,1}$ for
some $\ell^* \in \{1, \cd , \ka - 1\}$;

Case (2): $\jm^* = [S^{j}]_{\ell^*,1}$ for some $\ell^* \in \{1, \cd
, \ka\}$;

Case (3): $\jm^* > [S^{j}]_{\ka,1}$.

\bsk

In Case (1), define $\wh{S}^{j+1}$ as a $(\ka+1) \times 2$ matrix such that
\[
\begin{array}{ll}
[\wh{S}^{j+1}]_{\ell,1} = [S^{j}]_{\ell,1},
& \qqu [\wh{S}^{j+1}]_{\ell,2} = [S^{j}]_{\ell,2}, \qu \ell = 1, \cd,
\ell^*\\
{ [ \wh{S}^{j+1} ]} _{\ell^* + 1,1} = \jm^*,
& \qqu [\wh{S}^{j+1}]_{\ell^* + 1,2} = 1 + [S^{j}]_{\ell^*,2}\\
{[ \wh{S}^{j+1} ]} _{\ell + 1,1} = [S^{j}]_{\ell,1}, & \qqu
[\wh{S}^{j+1}]_{\ell + 1,2} = 1 + [S^{j}]_{\ell,2}, \qu \ell =
\ell^* + 1, \cd, \ka.
\end{array}
\]

In Case (2), define $\wh{S}^{j+1}$ as a $\ka \times 2$ matrix such that
\[
\begin{array}{ll}
{[\wh{S}^{j+1}]}_{\ell,1} = [S^{j}]_{\ell,1}, & \qqu
[\wh{S}^{j+1}]_{\ell,2} = [S^{j}]_{\ell,2}, \qu \ell = 1, \cd, \ell^* - 1\\
{[\wh{S}^{j+1}]}_{\ell,1} = [S^{j}]_{\ell,1}, & \qqu [\wh{S}^{j+1}]_{\ell,2}
= 1 + [S^{j}]_{\ell,2}, \qu \ell =
\ell^*, \cd, \ka.
\end{array}
\]

In Case (3), define $\wh{S}^{j+1}$ as a $(\ka+1) \times 2$ matrix such that
\[
\begin{array}{ll}
{[\wh{S}^{j+1}]}_{\ell,1} = [S^{j}]_{\ell,1}, & \qqu
[\wh{S}^{j+1}]_{\ell,2} = [S^{j}]_{\ell,2}, \qu \ell = 1, \cd, \ka\\
{[\wh{S}^{j+1}]}_{\ka + 1,1} = \jm^*, & \qqu [\wh{S}^{j+1}]_{\ka + 1,2} = 1
+ [S^{j}]_{\ka,2}.
\end{array}
\]

\bsk

\item [Step (3)] Let $\wh{\ka}$ denote the number of rows of
$\wh{S}^{j+1}$. If $[\wh{S}^{j+1}]_{\wh{\ka},2} < N$ then let
$S^{j+1} = \wh{S}^{j+1}$, otherwise find index $\ell_\star$ by a
bisection search such that $[\wh{S}^{j}]_{\ell_\star - 1,2} < N \leq
[\wh{S}^{j}]_{\ell_\star,2}$ and define $S^{j+1}$ as an $\ell_\star
\times 2$ matrix such that
\[
\begin{array}{ll}
{[S^{j+1}]}_{\ell, 1} = [\wh{S}^{j+1}]_{\ell,1},
& \qqu [S^{j+1}]_{\ell, 2} = [\wh{S}^{j+1}]_{\ell,2}, \qu \ell = 1, \cd,
\ell_\star - 1\\
{[S^{j+1}]}_{\ell_\star, 1} = [\wh{S}^{j+1}]_{\ell_\star,1}, & \qqu
[S^{j+1}]_{\ell_\star, 2} =
[\wh{S}^{j+1}]_{\ell_\star,2}.
\end{array}
\]

\eed \bsk

\subsubsection{Violations Tracking} \la{sec522}

In this section, we describe how to update the matrix of violations
in the case of $\bb{I}(X_{j+1}) = 1$. The key idea is to ensure
condition (\ref{con2}). Let $\ka$ be the number of rows of $S^j$.
Let $\tau$ be the number of rows of $V^j$. Let $\jm^*$ be the index
obtained in the process of updating $S^j$ such that $X_{j+1} \in
\mathcal{B}_ {r_i}$ for $\jm^* \leq i \leq m$ and $X_{j+1} \notin
\mathcal{B}_ {r_i}$ for $1 \leq i \leq \jm^* - 1$. We proceed as
follows.

\bed

\item [Step (i)] Identify the {\it maximal} index $\iota$ such
that the experiment for uncertainty radius $r_\iota$ has not been completed
by the following method.

\bsk

$\diamondsuit$ If $[S^j]_{\ka, 2} < N$, then let $\iota = \ka$, otherwise
find $\iota$ by a bisection search
such that $[V^j]_{\iota,1} < [S^j]_{\ka, 1}, \qu [V^j]_{\iota+1,1} \geq
[S^j]_{\ka, 1}$.

\bsk

\item [Step (ii)] Modify $V^j$ as a temporary matrix $\wh{V}^j$ based on
the following two cases.

\bsk

Case(a) : $[S^j]_{\ka, 2} < N$ or $[S^j]_{\ka, 2} = N, \qu
[V^j]_{\iota +1,1} = [S^j]_{\ka, 1}$.

Case(b) : $[S^j]_{\ka, 2} \geq N$ and the index $\iota$ guarantees
$[V^j]_{\iota+1,1} > [S^j]_{\ka, 1}$.

\bsk

In Case (a), we define $\wh{V}^j = V^j$. In Case (b), we define
$\wh{V}^j$ as a $(\tau + 1) \times 2 $ matrix such that
\[
\begin{array}{ll}
{[\wh{V}^{j}]}_{\ell,1} = [V^{j}]_{\ell,1}, & \qqu
[\wh{V}^{j}]_{\ell,2} = [V^{j}]_{\ell,2}, \qu \ell = 1, \cd, \iota\\
{[\wh{V}^{j}]}_{\iota+1,1} = [S^j]_{\ka, 1}, & \qqu [\wh{V}^{j}]_{\iota+1,2}
= [V^{j}]_{\iota,2}\\
{[\wh{V}^{j}]}_{\ell+1,1} = [V^{j}]_{\ell,1}, & \qqu [\wh{V}^{j}]_{\ell+1,2}
= [V^{j}]_{\ell,2}, \qu \ell =
\iota+1, \cd, \tau.
\end{array}
\]

\bsk

\item [Step (iii)] Obtain $V^{j+1}$ by modifying $\wh{V}^j$ based on the
following three cases.

\bsk

Case (i):  $[\wh{V}^{j}]_{\ell^*,1} < \jm^* < [\wh{V}^{j}]_{\ell^* +
1,1}$ for some $\ell^* \in \{1, \cd , \iota - 1\}$;

Case (ii):  $\jm^* = [\wh{V}^{j}]_{\ell^*,1}$ for some $\ell^* \in
\{1, \cd , \iota \}$;

Case (iii):  $\jm^* > [\wh{V}^{j}]_{\iota,1}$.

\bsk

Let $\wh{\tau}$ be the number of rows of $\wh{V}^{j}$. In Case (i),
define $V^{j+1}$ as a $(\wh{\tau} + 1) \times 2 $ matrix such that
\[
\begin{array}{ll}
{[V^{j+1}]}_{\ell,1} = [\wh{V}^{j}]_{\ell,1}, & \qqu
[V^{j+1}]_{\ell,2} = [\wh{V}^{j}]_{\ell,2}, \qu \ell = 1, \cd, \ell^*\\
{[V^{j+1}]}_{\ell^* + 1,1} = \jm^*, & \qqu
[V^{j+1}]_{\ell^* + 1,2} = 1 + [\wh{V}^{j}]_{\ell^*,2}\\
{[V^{j+1}]}_{\ell + 1,1} = [\wh{V}^{j}]_{\ell,1}, & \qqu
[V^{j+1}]_{\ell + 1,2} = 1 + [\wh{V}^{j}]_{\ell,2}, \qu \ell =
\ell^* + 1,
\cd, \iota\\
{[V^{j+1}]}_{\ell + 1,1} = [\wh{V}^{j}]_{\ell,1}, & \qqu [V^{j+1}]_{\ell +
1,2} = [\wh{V}^{j}]_{\ell,2}, \qu
\ell = \iota + 1, \cd, \wh{\tau}.
\end{array}
\]

In Case (ii), define $V^{j+1}$ as a $\wh{\tau} \times 2 $ matrix
such that
\[
\begin{array}{ll}
{[V^{j+1}]}_{\ell,1} = [\wh{V}^{j}]_{\ell,1}, & \qqu
[V^{j+1}]_{\ell,2} = [\wh{V}^{j}]_{\ell,2}, \qu \ell = 1, \cd, \ell^* - 1\\
{[V^{j+1}]}_{\ell,1} = [\wh{V}^{j}]_{\ell,1}, & \qqu
[V^{j+1}]_{\ell,2} = 1 + [\wh{V}^{j}]_{\ell,2}, \qu \ell = \ell^*,
\cd,
\iota\\
{[V^{j+1}]}_{\ell,1} = [\wh{V}^{j}]_{\ell,1}, & \qqu [V^{j+1}]_{\ell,2} =
[\wh{V}^{j}]_{\ell,2}, \qu \ell =
\iota + 1, \cd, \wh{\tau}.
\end{array}
\]

In Case (iii), define $V^{j+1}$ as a $(\wh{\tau} + 1) \times 2 $ matrix such
that
\[
\begin{array}{ll}
{[V^{j+1}]}_{\ell,1} = [\wh{V}^{j}]_{\ell,1}, & \qqu
[V^{j+1}]_{\ell,2} = [\wh{V}^{j}]_{\ell,2}, \qu \ell = 1, \cd, \iota\\
{[V^{j+1}]}_{\iota + 1,1} = \jm^*, & \qqu
[V^{j+1}]_{\iota + 1,2} = 1 + [\wh{V}^{j}]_{\iota,2}\\
{[V^{j+1}]}_{\ell + 1,1} = [\wh{V}^{j}]_{\ell,1}, & \qqu [V^{j+1}]_{\ell +
1,2} = [\wh{V}^{j}]_{\ell,2}, \qu
\ell = \iota + 1, \cd, \wh{\tau}.
\end{array}
\]

\eed

\subsection{Complexity of Data Processing and Memory}

It can be seen that the memory requirement and the computation due to data
processing are determined by the
sizes of matrices $S^j$ and $V^j$. To quantify the complexity, we have the
following results.

\beT \la{Memory_bound}

For any $j$, the following statements hold true:

\bed

\item [(I)]
The number of rows of matrix $S^j$ is no more than $N$;

\item [(II)] The expected number of rows of matrix $V^j$ is no
greater than \be \la{ieq_encp} 1 + N \li [ P_e (a) + 2 d
\int_{\f{a}{\hbar} }^a \f{ P_e (x) } { x } d x \ri ] \leq 1 + N P_e
(a) \li ( 1 + 2 d \ln \hbar \ri ) \ee where $P_e (x) = 1 - \min_{y
\in [\f{a}{\lm}, \; x]} \bb{P}(y), \; \fa x \in \li [\f{a}{\lm}, \;
a \ri ]$ and
\[
\hbar = \max \li ( \min \li (\lm, \f{a}{\rho_0} \ri ), \; 1 \ri )
\]
with $\ro_0 = \sup \{r \mid \bb{P}(r) = 1 \}$. \eed

\eeT

\bsk

See Appendix F for a proof.  We now revisit the robustness analysis problem discussed in Remark 1 of Section 4
from the perspective of memory complexity. Assume that each data record $(k_i, n_i)$ (or each row of $V^j$)
occupies $4$ bytes of computer internal memory (RAM). As illustrated in Section $5.2$, when using the
conventional data structure, it takes $13$G (giga bytes) of RAM to save the data and visualize the results. On
the other hand, in our new algorithm, if the smallest proportion is $p_* = \min_{ r \in \li [ \f{a}{\lm}, a \ri
] } \bb{P} (r) > 0.999$ and $\hbar < \f{3}{2}$, the RAM requirement will be equivalent to \bee
&  & 4 \times \li [ 1 + ( 1- p_*) \li ( 1 + 2 d \ln \hbar \ri ) N \ri ]\\
& = & 4 \times \li [ 1 + (1 - 0.999) \times
\li ( 1 + 2 \times 1800 \times \ln \f{3}{2}  \ri ) \times 10^6 \ri ]\\
& < & 6.2 \times 10^6 \; \tx{bytes} \ap 6.2 \; \tx{M} \; \tx{bytes}.
\eee It can be seen that such requirement of memory is extremely low
as compared to that of the conventional method.  Theorem
\ref{Memory_bound} also reveals that the complexity of data
processing is very low.

\subsection{Confidence Band}

To be useful, any numerical techniques should provide a method for
error assessment. Monte Carlo simulation is no exception. The
following results allows us to construct confidence band for the
robustness curve. Such post-experimental statistical inference can
remedy the conservatism of {\it a priori} choice of sample size $N$
based on the Chernoff bound. In order to overcome the computational
complexity of the Clopper-Pearson's confidence interval \cite{Clo},
we have developed new methods to facilitate the construction of the
confidence band.

\beT \la{CI_band} Let $\de \in (0,1)$. Let $\mcal{ L} (k) =
\frac{k}{N} + \frac{3}{4} \; \frac{ 1 - \frac{2k}{N} - \sqrt{ 1 + 4
\theta \; k ( 1- \frac{k}{N}) } } {1 + \theta N }$ and $\mcal{ U}
(k) = \frac{k}{N} + \frac{3}{4} \; \frac{ 1 - \frac{2k}{N} + \sqrt{
1 + 4 \theta \; k ( 1- \frac{k}{N}) } } {1 + \theta N }$ with
$\theta = \frac{9}{ 8 \ln \frac{2}{\delta} }$. Let $\ze = \f{ r -
r_i } {r_{i+1} - r_i }$. Let \[ K_i = N - \bs{v}(i, \bs{n}), \qqu i
= 1, \cd, m.
\]
Let $\varsigma = 1 - \f{g(r_\star) } { r_{i+1} - r_i }$. Define
$\overline{\bb{P}} (r) = \ze \; \mathcal{U} (K_i) + (1 - \ze) \;
\mathcal{U} (K_{i+1}) + \varsigma$ and $\underline{\bb{P}} (r) = \ze
\; \mathcal{L} (K_i) + (1 - \ze) \; \mathcal{L} (K_{i+1}) -
\varsigma$. Then
\[ \Pr \{ \underline{\bb{P}} (r) < \bb{P}(r) <
\overline{\bb{P}} (r), \; \fa r \in [r_i, r_{i+1}] \} > 1 - \de.
\] \eeT

See Appendix G for a proof. The family of intervals
$[\underline{\bb{P}} (r), \; \overline{\bb{P}} (r)], \; r \in [ a
\sh \lm, \; a]$ is referred to as the {\it confidence band}. It is
important to note that the confidence band can be efficiently
constructed by making use of the piece-wise constant property of
$\bs{v}(i, \bs{n})$. It can be shown that the computational
complexity of constructing the confidence band is also absolutely
bounded.

\section{Conclusion}

It is possible to make a case for the statement that the
probabilistic robustness analysis is essentially the study of the
robustness function, especially about its probabilistic
implications, efficient evaluation and computational complexity. We
have addressed these issues in this paper. In particular, we have
developed randomized algorithms which offer more insights for system
robustness. We rigorously show that, in both aspects of computer
running time and memory requirement, the complexity of such
randomized algorithms is not only linear in the dimension of
uncertainty space, but also surprisingly low. While the complexity
of conventional method grows linearly with the number of grid points
and the error due to interpolation is not well controlled, our
techniques completely resolve such issues. In short, our method
guarantees accuracy and efficiency.

\appendix

\section{Proof of Theorem \ref{Bound_General}}

We first establish a basic inequality that will be used to prove the
theorem. \begin{lemma} \label{inc} For any
$x
>1$,
\[
\frac{1}{x} + \ln x > 1.
\]
\end{lemma}

{\it Proof.} Let \[ f(x) = \frac{1}{x} + \ln x.
\]
Then $f(1) = 1$ and
\[
\frac{d \; f(x) } {d x } = \frac{x - 1}{x^2} > 0, \qquad \forall x >
1.
\]
It follows that $f(x) > 1, \; \forall x > 1$. \hfill\mbox{$\Box$}

Now we are in the position to prove Theorem \ref{Bound_General}.
Observing that
\[
\li ( \frac{r_m}{r_1} \ri )^d = \prod_{i=1}^{m-1} \li ( \frac{r_{i+1}}{r_i}
\ri )^d,
\]
we have \[ \ln \left( \frac{r_m}{r_1} \right )^d = \sum_{i=1}^{m-1} \ln
\left ( \frac{r_{i+1}}{r_i} \right )^d.
\]
Therefore, \begin{eqnarray*} \sum_{i=1}^{m-1} \li(
\frac{r_i}{r_{i+1}} \ri )^d + \ln \left( \frac{r_m}{r_1} \right )^d
& = & \sum_{i=1}^{m-1} \left [ \frac{1 } { \li ( \frac{r_{i+1}}{r_i}
\ri )^d} + \ln \left ( \frac{r_{i+1}}{r_i} \right )^d \right ].
\end{eqnarray*} Since $\li( \frac{r_{i+1}}{r_i} \ri )^d > 1, \; i = 1,
\cdots, m-1$, it follows from Lemma \ref{inc} that
\[
\frac{1 } { \li ( \frac{r_{i+1}}{r_i} \ri )^d } + \ln \left (
\frac{r_{i+1}}{r_i} \right )^d > 1, \qquad i = 1, \cdots , m-1.
\]
Hence,
\[
\sum_{i=1}^{m-1} \li ( \frac{r_i}{r_{i+1}} \ri )^d + \ln \left(
\frac{r_m}{r_1} \right )^d
>m - 1,
\]
or equivalently,
\[
m - \sum_{i=1}^{m-1} \li ( \frac{r_i}{r_{i+1}} \ri )^d < 1 + \ln
\left( \frac{r_m}{r_1} \right )^d = 1 + d \ln \lm.
\]

Finally, by Theorem 1 of \cite{C0} and the definition of
$m_{\mrm{eq}}$, we have \[ m_{\mrm{eq}} = m - \sum_{i=1}^{m-1} \li (
\frac{r_i}{r_{i+1}} \ri )^d < 1 + d \ln \lm.
\]

\section{Proof of Theorem \ref{bound1}}

To prove the theorem, we need some preliminary results. It is
derived in \cite{BLT} that $\li | \f{d \bb{P}(r) }{ d r} \ri | <
\f{2 d}{r}$ when $\bb{P}(.)$ is differentiable. The following lemma
indicates that the bound on the rate of variation of $\bb{P}(.)$ can
be much tighter.

\beL \label{boundness} For arbitrary robustness requirement,
\[
\li | \bb{P}(r + \vDe r) - \bb{P}(r) \ri | \leq 1 - \li ( 1 + \f{
\vDe r } { r } \ri )^{-d} < \f{d}{r} \; \vDe r
\]
for any $r > 0$ and any $\vDe r > 0$. \eeL

\bpf Let $\mcal{Q}_r \leu \mcal{B}_r$ be the set such that the
robustness requirement is satisfied. Let
\[
I_1 = \f{ \tx{vol} ( \mcal{Q}_{r + \vDe r} ) } { \tx{vol} ( \mcal{B}_{r +
\vDe r} ) } - \f{ \tx{vol} (
\mcal{Q}_{r} ) } { \tx{vol} ( \mcal{B}_{r + \vDe r} ) }, \qqu I_2 = \f{
\tx{vol} ( \mcal{Q}_{r} ) } { \tx{vol} (
\mcal{B}_{r + \vDe r} ) } - \f{ \tx{vol} ( \mcal{Q}_{r} ) } { \tx{vol} (
\mcal{B}_{r} ) }.
\]
Let ``$\setminus$'' denote the operation of set minus. Observing
that $ \mcal{Q}_{r + \vDe r} \setminus \mcal{Q}_{r} \subseteq
\mcal{B}_{r + \vDe r} \setminus \mcal{B}_{r}$, we have $\tx{vol} (
\mcal{Q}_{r + \vDe r} ) - \tx{vol} ( \mcal{Q}_{r} ) \leq \tx{vol} (
\mcal{B}_{r + \vDe r} ) - \tx{vol} ( \mcal{B}_{r} )$. Using this
fact and the identity $\tx{vol} ( \mcal{B}_r ) = r^d \; \tx{vol} (
\mcal{B}_1 )$, we have
\[
0 \leq I_1 \leq \f{ \tx{vol} ( \mcal{B}_{r + \vDe r} ) - \tx{vol} (
\mcal{B}_{r} ) } { \tx{vol} ( \mcal{B}_{r +
\vDe r} ) } = 1 - \li ( 1 + \f{ \vDe r } { r } \ri )^{-d}
\]
and
\[
- \li[ 1 - \li ( 1 + \f{ \vDe r } { r } \ri )^{-d} \ri ] \leq - \f{ \tx{vol}
( \mcal{Q}_{r} ) } { \tx{vol} (
\mcal{B}_{r} ) } \f{ \tx{vol} ( \mcal{B}_{r + \vDe r} ) - \tx{vol} (
\mcal{B}_{r} ) } { \tx{vol} ( \mcal{B}_{r +
\vDe r} ) } = I_2 \leq 0.
\]
Therefore, $\li | \bb{P}(r + \vDe r) - \bb{P}(r) \ri | = |I_1 + I_2|
\leq 1 - \li ( 1 + \f{ \vDe r } { r } \ri )^{-d} < \f{d}{r} \vDe r$
where the last inequality follows from inequality $1 - \li ( 1 + \f{
x } { r } \ri )^{-d} < \f{d x}{r}, \; \fa x > 0$. To prove this
inequality, we can define function $ h(x) \DEF 1 - \li ( 1 + \f{ x }
{ r } \ri )^{-d} - \f{d x}{r}, \; x > 0$ and check that $h(0) = 0$
and $\f{\pa h(x) } { \pa x } = \f{d}{r} \li [ \li ( 1 + \f{ x } { r
} \ri )^{-(d+1)} - 1 \ri] < 0, \; \fa x > 0$.

\epf

We are now in the position to prove the theorem. It can be shown
that
\begin{eqnarray} | \bb{P} (r) - \bb{P}^* (r) | & = & \li | \f{
(r_{i+1} - r) [\bb{P} (r) - \bb{P} (r_i) ] + (r - r_{i}) [\bb{P}
(r) - \bb{P} (r_{i+1}) ] } { r_{i+1} - r_i } \ri | \nonumber\\
& \leq & \f{ (r_{i+1} - r) |\bb{P} (r) - \bb{P} (r_i) | + (r -
r_{i}) |\bb{P} (r) - \bb{P} (r_{i+1}) | } { r_{i+1} - r_i }
\la{eqa}.
\end{eqnarray}
By Lemma~\ref{boundness} and inequality (\ref{eqa}), we have \bee |
\bb{P} (r) - \bb{P}^* (r) | & \leq & \f{ (r_{i+1} - r) \li [ 1 - \li
( \f{r}{r_i} \ri )^{-d} \ri ] + (r - r_{i}) \li [ 1 - \li (
\f{r_{i+1}}{r} \ri )^{-d} \ri ] } { r_{i+1} - r_i
}\\
& = & 1 - \f{ g(r) } { r_{i+1} - r_i }. \eee Note that
\[
\f{\pa g(r)}{\pa r} = \Phi (r) - \Psi(r)
\]
where
\[
\Phi (r) = \li ( \f{r_{i+1}}{r} \ri )^{-d} \li [ 1 + \li ( 1 -
\f{r_i} { r } \ri ) d \ri ], \qqu \Psi(r) = \li ( \f{r}{r_{i}} \ri
)^{-d} \li [ 1 + \li ( \f{r_{i+1}} { r } - 1 \ri ) d \ri ].
\]
It can be verified that
\[
\Phi (r_i) = \li ( \f{r_{i+1}}{r_i} \ri )^{-d} < 1, \qqu \Psi(r_i) =
1 + \li ( \f{r_{i+1}} { r_i } - 1 \ri ) d > 1, \qqu \li . \f{\pa
g(r)}{\pa r} \ri |_{ r = r_i} < 0,
\]
\[
\Phi (r_{i+1}) = 1 + \li ( 1 - \f{r_{i}} { r_{i+1} } \ri ) d > 1,
\qu \Psi(r_{i+1}) = \li ( \f{r_{i+1}}{r_i} \ri )^{-d} < 1, \qu \li .
\f{\pa g(r)}{\pa r} \ri |_{ r = r_{i+1}} > 0.
\]
It can be checked that $\Phi(r)$ is a monotone increasing function
of $r$ and that $\Psi(r)$ is a monotone decreasing function of $r$.
Hence, $\f{\pa g(r)}{\pa r}$ is a monotone increasing function of
$r$. Moreover, there exists an unique $r_\star \in (r_i, \;
r_{i+1})$ such that $\li . \f{\pa g(r)}{\pa r} \ri |_{ r = r_\star}
= 0$, i.e., $\Phi (r_\star) = \Psi (r_\star)$. Furthermore, $g(r)$
is a convex function of $r$. Consequently,
\[
\min_{ r \in [r_i, \; r_{i+1}]} g(r) = g(r_\star)
\]
and we have shown
\[
| \bb{P} (r) - \bb{P}^* (r) | \leq 1 - \f{ g(r_\star) } { r_{i+1}
- r_i } \qqu \fa r \in [r_i, \; r_{i+1}].
\]
Since $\f{\pa g(r)}{\pa r}$ is a monotone increasing function of
$r$, we can compute $r_\star$ by a bisection search over interval
$(r_i, \; r_{i+1})$.

By Lemma~\ref{boundness} and inequality (\ref{eqa}), we have \bee |
\bb{P} (r) - \bb{P}^* (r) | & \leq & \f{ (r_{i+1} - r) (r - r_i)
\f{d} {r_i}  + (r - r_{i})
(r_{i+1} - r) \f{d} {r} } { r_{i+1} - r_i }\\
& \leq & \f{ (r_{i+1} - r) (r - r_i) \f{d} {r_i}  + (r - r_{i})
(r_{i+1} - r) \f{d} {r_i} } { r_{i+1} - r_i }\\
& \leq & \f{2 d} {r_i (r_{i+1} - r_i)} \; \max_{r \in [r_i,
r_{i+1}]} (r_{i+1} - r) (r - r_i)\\
& = & \f{2 d} {r_i (r_{i+1} - r_i)} \; \f{ (r_{i+1} - r_i)^2 }
{4}\\
& = & \f{ d (r_{i+1} - r_i) } {2 r_i}. \eee

\section{Proof of Theorem \ref{grid_uni}}

By Theorem \ref{bound1}, $|\bb{P} ( r ) - \bb{P}^* ( r)| \leq \f{ d
\; (r_{i+1} - r_i) } {2 r_i}, \qu \fa r \in [r_i, r_{i+1}]$. Thus,
it suffices to show $\f{ d \; (r_{i+1} - r_i) } {2 r_i} < \ep$,
i.e., \be \la{con3} \f{ r_{i+1} }{ r_i } < 1 + \f{2 \ep}{d}. \ee By
definition (\ref{def2}), for $i = 1, \cd, m-1$, \bee \f{ r_{i+1} }
{r_i} & = & \f{ a - \f{ (m - i - 1) (\lm - 1) }
{ (m -1) \lm } a } { a - \f{ (m - i) (\lm - 1) }{ (m -1) \lm } a }\\
& = & 1 + \f{ \lm - 1 } { m- 1 + (\lm -1) (i -1) }\\
& \leq & 1 + \f{ \lm - 1 } { m - 1 }. \eee By virtue of
(\ref{con3}), to guarantee that the gridding error is less than
$\ep$, it suffices to ensure $1 + \f{ \lm - 1 } { m - 1 } < 1 + \f{2
\ep}{d}$, i.e., $m > 1 + \f{d (\lm -1)}{2 \ep}$. Hence, it suffices
to have
\[
m \geq 2 + \li \lf \f{ (\lm -1) d } { 2 \ep } \ri \rf.
\]
It can be verified that
\[
\f{ r_i } { r_{i+1} } = 1 - \f{ 1 } { \f{m-1} {\lm -1} + i }, \qqu i
= 1, \cd, m-1.
\]
By Theorem 1 of \cite{C0}, the sample reuse factor is given by \bee \mcal{
F}_\mathrm{reuse} & = & \frac{m} {m -
\sum_{i=1}^{m-1} \left(
\frac{r_i}{r_{i+1}} \right)^d}\\
& = & \f{m} { m - \sum_{i=1}^{m-1} \li ( 1 - \f{1} { \f{m-1} {\lm-1}
+ i } \ri )^d }. \eee Therefore,
\[
m_{\mrm{eq}} (\epsilon) = \f{m} { \mcal{ F}_\mathrm{reuse} } = m -
\sum_{i=1}^{m-1} \li ( 1 - \f{1} { \f{m-1} {\lm-1} + i } \ri )^d.
\]

\section{Proof of Theorem \ref{Grid_geometric}}

By virtue of (\ref{def}), we have $\f{ r_{i+1} }{ r_i } = \lm^{
\f{1}{m-1} }$. Hence, by (\ref{con3}), it suffices to show $\lm^{
\f{1}{m-1} } < 1 + \f{2 \ep}{d}$, which can be reduced to $m > 1 +
\f{ \ln \lm } { \ln \li ( 1 + \f{2 \ep}{d} \ri ) }$. This inequality
is equivalent to $m \geq 2 + \li \lf \f{ \ln \lm } { \ln \li ( 1 +
\f{2 \ep}{d} \ri ) } \ri \rf$. By equation (\ref{eqreuse}) and
Theorem 1 of \cite{C0}, we have $\bb{E} [ \bs{n} ] = \li [ m - (m-1)
\left( \frac{1}{\lm} \right)^{ \f{d}{m -1} } \ri ] N$ and hence
obtain $m_{\mrm{eq}} (\ep)$.  Note that \[ \mcal{ F}_\mathrm{reuse}
= \f{ m } { m_{\mrm{eq}} (\epsilon) }
 > \f{ m } { 1 + d \; \ln \lm  }
= \f{ 2 + \li \lf \f{ \ln \lm } { \ln \li ( 1 + \f{2 \ep}{d} \ri ) }
\ri \rf } { 1 + d \; \ln \lm } >  \f{ \f{ \ln \lm } { \ln \li ( 1 +
\f{2 \ep}{d} \ri ) } } { 1 + d \; \ln \lm  }. \] Making use of the
inequality $\ln (1 + x) < x, \; \fa x > 0$, we have $\ln \li ( 1 +
\f{2 \ep}{d} \ri ) < \f{2 \ep}{d}$. Therefore, \[ \mcal{
F}_\mathrm{reuse} > \f{ \f{ \ln \lm }
{ \f{2 \ep}{d} }  } { 1 + d \; \ln \lm  }\\
= \f{1}{2 \ep} \li ( 1 - \f{1}{ 1 + d \; \ln \lm } \ri ). \]

\bsk

\section{Proof of Theorem \ref{Memory_bound}}

\bed

\item [Proof of statement (I)] Obviously, $[S^j]_{1,2} \geq 1, \qu
[S^j]_{\ka,2} \leq N$.  From the rules of sampling, we can perform
induction with respect to $j$ and have $[S^j]_{\ell+1,2} -
[S^j]_{\ell,2} \geq 1, \; \ell = 1, \cd, \ka - 1$. Observing that
\bee [S^j]_{\ka,2} & = & [S^j]_{1,2} +
\sum_{\ell = 1}^{\ka - 1} \li ( [S^j]_{\ell+1,2} - [S^j]_{\ell,2} \ri )\\
& \geq & 1 + \sum_{\ell = 1}^{\ka - 1}
\li ( [S^j]_{\ell+1,2} - [S^j]_{\ell,2} \ri )\\
& \geq & 1 + \ka - 1\\
& = & \ka, \eee we have $\ka \leq [S^j]_{\ka,2} \leq N$.

\item [Proof of statement (II)] We need some preliminary results.

\beL \la{lem2} Let $1 \leq i \leq m - 1$. Then
\[
\li | \li [ 1 - \li ( \f{ r_i } { r_{i+1} } \ri )^d \ri ] - \f { d
(r_{i+1} - r_i) } { r_{i+1} } \ri | \leq \f{ d(d-1) } { 2 } \; \li (
\f { r_{i+1} - r_i } { r_{i+1} } \ri )^2.
\]
\eeL

\bpf Note that \bee \li | \li [ 1 - \li ( \f{ r_i } { r_{i+1} } \ri
)^d \ri ] - \f { d (r_{i+1} - r_i) } { r_{i+1} } \ri | & = & \li |
\li ( \f{ r_i } { r_{i+1} } \ri )^d -
\li ( 1 - \f { d (r_{i+1} - r_i) } { r_{i+1} } \ri ) \ri |\\
& = & \li | \li ( 1 - t \ri )^d - \li ( 1 - d \; t \ri ) \ri | \eee
where $t = \f { r_{i+1} - r_i } { r_{i+1} }$. It can be checked that
$\li | \li ( 1 - t \ri )^d - \li ( 1 - d \; t \ri ) \ri | = \f{
d(d-1) } { 2 } t^2$ for $d = 1, 2$. For $d \geq 3$, by Taylor's
expansion formula, there exists $\xi \in (0, t)$ such that
\[
(1 - t )^d = 1 - d \; t + \f{ d(d-1) } { 2 } ( 1- \xi)^{d-2} \; t^2.
\]
Observing that $0 < t < 1$ since $0 < r_i < r_{i+1}$, we hence have
$0 < ( 1- \xi)^{d-2} < 1$ and $\li | \li ( 1 - t \ri )^d - \li ( 1 -
d \; t \ri ) \ri | < \f{ d(d-1) } { 2 } t^2$ for $d \geq 3$.
Therefore, for any $d \geq 1$,
\[
\li | \li [ 1 - \li ( \f{ r_i } { r_{i+1} } \ri )^d \ri ] - \f { d
(r_{i+1} - r_i) } { r_{i+1} } \ri | \leq \f{ d(d-1) } { 2 } t^2 =
\f{ d(d-1) } { 2 } \; \li ( \f { r_{i+1} - r_i } { r_{i+1} } \ri
)^2.
\]

\epf

\beL \la{lem22} Define the maximum gap between grid points as
\[
\varpi = \max_{1 \leq i \leq m-1} \; (r_{i+1} - r_i). \] Then
\[
\sum_{i=1}^{m-1} \li ( \f { r_{i+1} - r_i } { r_{i+1} } \ri )^2 <
\f{ \lm (\lm -1) \varpi } { a }.
\]
\eeL

\bpf

Note that \[ \sum_{i=1}^{m-1} \li ( \f { r_{i+1} - r_i } { r_{i+1} }
\ri )^2 \leq \varpi \sum_{i=1}^{m-1} \f { r_{i+1} - r_i } {
r_{i+1}^2 } < \varpi \sum_{i=1}^{m-1} \f { r_{i+1} - r_i } { \li(
\f{a}{\lm} \ri )^2 }.
\]
By successive cancelation, \[ \sum_{i=1}^{m-1} (r_{i+1} - r_i) = r_m
- r_1 = a - \f{a}{\lm}. \] Hence, \bee \sum_{i=1}^{m-1} \li ( \f {
r_{i+1} - r_i } { r_{i+1} } \ri )^2 < \varpi \sum_{i=1}^{m-1} \f {
r_{i+1} - r_i } { \li( \f{a}{\lm} \ri )^2 } =  \varpi \; \f { a -
\f{a}{\lm} } { \li( \f{a}{\lm} \ri )^2 } = \f{ \lm (\lm -1) \varpi }
{ a }. \eee

\epf

\beL \la{lem889} The expected number of rows of the matrix of
violations $V^{\bs{n}}$ is no greater than $1 + N P_e(a) + 2 N
\sum_{j=1}^{m-1} P_e(r_j) \li [ 1 - \li ( \f{r_j}{r_{j+1}} \ri )^d
\ri ]$. \eeL

\bpf Let $X_1^j, \cd, X_{ \mathbf{n}_j }^j$ be the samples generated
from uncertainty set with radius $r_j$. Let $Y_i^j = \bb{I} (X_i^j),
\qu i = 1, \cd, \mathbf{n}_j$. By the principle of sample reuse, the
value of $\mathbf{n}_j$ depends only on the samples generated from
uncertainty sets with radius $r_k, \; j + 1 \leq k \leq m$.
Consequently, event $\{\mathbf{n}_j = \nu \}$ is independent of
event $\{ Y_i^j = 1 \}$ and $\Pr \{ Y_i^j = 1, \; \mathbf{n}_j = \nu
\} = \Pr \{ Y_i^j = 1 \} \Pr \{ \mathbf{n}_j = \nu \}$. By the
definitions of $Y_i^j$ and $P_e(.)$, we have $\Pr \{ Y_i^j = 1 \} =
1 - \bb{P} (r_j) \leq P_e(r_j)$.  Therefore, \bee \bb{E} \li [
\sum_{i=1}^{\mathbf{n}_j} Y_i^j \ri ] & = & \sum_{\nu=1}^{N}
\sum_{i=1}^\nu \Pr \{ Y_i^j = 1, \;
\mathbf{n}_j = \nu \}\\
& = & \sum_{\nu=1}^{N} \sum_{i=1}^\nu \Pr \{ Y_i^j = 1 \} \Pr \{
\mathbf{n}_j = \nu \}\\
& \leq & \sum_{\nu=1}^{N} \nu \; P_e (r_j) \Pr \{ \mathbf{n}_j = \nu \}\\
& = & P_e (r_j) \; \bb{E} [\mathbf{n}_j]\eee for $j = 1, \cd, m$. We
now consider $V^{\bs{n}}$ with $\bs{n} = \sum_{i=1}^m \mathbf{n}_i$.
By the mechanism of the sample reuse algorithms, for $j=1, \cd,
m-1$, every new sample from uncertainty set with radius $r_j$ at
most creates $2 Y_i^j, \; i = 1, \cd, \mathbf{n}_j$ new rows for the
matrix of violations (see Section $6.2.2$). Note that $X_1$ create
at most $1 + Y_1^m$ rows for $V^1$. Every new sample from
uncertainty set with radius $r_m$ at most creates $Y_i^m, \; i = 2,
\cd, \mathbf{n}_m$ new rows for the matrix of violations. Hence
\begin{eqnarray}
&  & \bb{E} [\tx{The number of rows of matrix $V^{\bs{n}}$} ] \nonumber\\
& \leq & 1 + \bb{E} \li [ \sum_{i=1}^{\mathbf{n}_m} Y_i^m \ri ] + 2
\; \bb{E} \li [ \sum_{j=1}^{m-1} \sum_{i=1}^{\mathbf{n}_j} Y_i^j \ri
]
\nonumber\\
& \leq & 1 + P_e (r_m) \; \bb{E} [ \mathbf{n}_m ] + 2
\sum_{j=1}^{m-1} P_e (r_j) \; \bb{E} [ \mathbf{n}_j ] \nonumber.
\end{eqnarray} By Lemma 6 of \cite{C0}, we have \be \la{ineqb}
\bb{E} [ \mathbf{n}_j ] = \li [ 1 - \li ( \f{r_j}{r_{j+1}} \ri )^d
\ri ] N, \qqu j = 1, \cd, m-1. \ee By (\ref{ineqb}) and using the
fact that $\bb{E} [ \mathbf{n}_m ] = N, \; P_e (r_m) = P_e (a)$, we
have
\[
\bb{E} [\tx{The number of rows of matrix $V^{\bs{n}}$} ] \leq 1 + N
P_e(a) + 2 N \sum_{j=1}^{m-1} P_e(r_j) \li [ 1 - \li (
\f{r_j}{r_{j+1}} \ri )^d \ri ].
\]

\epf

\beL \la{lem99} For any grid scheme, \bee &  & \li |
\sum_{i=1}^{m-1} P_e(r_i) \li [ 1 - \li ( \f{ r_i } { r_{i+1} } \ri
)^d \ri ] - d
\sum_{i=1}^{m-1} \f { P_e(r_{i+1})(r_{i+1} - r_i ) } { r_{i+1} } \ri | \\
& < & \f{ d(d-1) \lm (\lm -1) \varpi} { 2a } + \f{ \lm d } { a} \;
\sum_{i=1}^{m-1} P_e(r_{i+1})(r_{i+1} - r_i ) - \f{ \lm d } { a} \;
\sum_{i=1}^{m-1} P_e(r_i)(r_{i+1} - r_i ). \eee \eeL

\bpf Note that \bee &  & \li | \sum_{i=1}^{m-1} P_e(r_i) \li [ 1 -
\li ( \f{ r_i } { r_{i+1} } \ri )^d \ri ] - d
\sum_{i=1}^{m-1} \f { P_e(r_{i+1})(r_{i+1} - r_i ) } { r_{i+1} } \ri |\\
& \leq & \sum_{i=1}^{m-1} \li |  P_e(r_i) \li [ 1 - \li ( \f{ r_i }
{ r_{i+1} } \ri )^d \ri ] -
d \f { P_e(r_i)(r_{i+1} - r_i ) } { r_{i+1} } \ri |\\
&  & + \sum_{i=1}^{m-1} \li | d \f { P_e(r_i)(r_{i+1} - r_i ) }
{ r_{i+1} } - d \f { P_e(r_{i+1})(r_{i+1} - r_i ) } { r_{i+1} } \ri |\\
& < & \sum_{i=1}^{m-1} \li | \li [ 1 - \li ( \f{ r_i } { r_{i+1} }
\ri )^d \ri ] -
\f { d (r_{i+1} - r_i ) } { r_{i+1} } \ri |\\
&  & + \f{ \lm d } { a} \sum_{i=1}^{m-1} \li [ P_e(r_{i+1})(r_{i+1}
- r_i ) - P_e(r_i)(r_{i+1} - r_i ) \ri ] \eee where the last
inequality follows from the facts that $0 \leq P_e(r_i) \leq
P_e(r_{i+1}) \leq 1$ and $r_{i+1} > \f{a}{\lm}$. Making use of Lemma
\ref{lem2} and Lemma \ref{lem22}, we have \bee &  & \li |
\sum_{i=1}^{m-1} P_e(r_i) \li [ 1 - \li ( \f{ r_i } { r_{i+1} } \ri
)^d \ri ] - d
\sum_{i=1}^{m-1} \f { P_e(r_{i+1})(r_{i+1} - r_i ) } { r_{i+1} } \ri |\\
& \leq & \f{ d(d-1) } { 2 } \; \sum_{i=1}^{m-1} \li ( \f { r_{i+1} -
r_i }
{ r_{i+1} } \ri )^2 \\
&  & + \f{ \lm d } { a} \; \sum_{i=1}^{m-1} P_e(r_{i+1})(r_{i+1} -
r_i )
- \f{ \lm d } { a} \; \sum_{i=1}^{m-1} P_e(r_i)(r_{i+1} - r_i )\\
& \leq & \f{ d(d-1) \lm (\lm -1) \varpi} { 2a } + \f{ \lm d } { a}
\; \sum_{i=1}^{m-1} P_e(r_{i+1})(r_{i+1} - r_i ) - \f{ \lm d } { a}
\; \sum_{i=1}^{m-1} P_e(r_i)(r_{i+1} - r_i ). \eee

\epf

\beL \la{Uencp} For a set of grid points $\mathcal{G} = \{ r_\ell
\mid 1 \leq \ell \leq m \}$ with $\f{a}{\lm} = r_1 < r_2 < \cd <
r_{m} = a$, define function $\aleph(.)$ such that
\[
\aleph ( \mathcal{G} ) = \sum_{\ell=1}^{m-1} P_e(r_\ell) \li [ 1 -
\li ( \f{r_\ell}{r_{\ell+1}} \ri )^d \ri
].
\]
Then for any two sets of grid points $\mathcal{G}_1$ and
$\mathcal{G}_2$ such that $\mathcal{G}_1 \subset \mathcal{G}_2$,
\[
\aleph ( \mathcal{G}_1 ) \leq \aleph ( \mathcal{G}_2 ).
\]
\eeL

\bpf Consider two sequences of grid points $\mathcal{G}_1 = \{
r_\ell \mid 1 \leq \ell \leq m\}$ and $\mathcal{G}_2 = \{
\wh{r}_\ell \mid 1 \leq \ell \leq m + 1 \}$ such that
\[
\f{a}{\lm} = r_1 < r_2 < \cd < r_{m} = a, \qqu \f{a}{\lm} = \wh{r}_1
< \wh{r}_2 < \cd < \wh{r}_{m+1} = a, \] and that $\mathcal{G}_2$ is
obtained from $\mathcal{G}_1$ by adding a grid point $\wh{r}_{i+1}$
to interval $(r_{i}, r_{i+1})$ where $1 \leq i \leq m-1$, i.e.,
$\wh{r}_j = r_j, \; j = 1, \cd, i$ and $\wh{r}_{j+1} = r_j, \; j =
i+1, \cd, m$. By the definition of function $\aleph (.)$, we have
\bee &  & \aleph (\mathcal{G}_2) - \aleph (\mathcal{G}_1)\\
& = & \sum_{\tau=1}^m P_e
( \wh{r}_\tau ) \li [ 1 - \li ( \f { \wh{r}_\tau } { \wh{r}_{\tau+1}
} \ri )^d \ri ] - \sum_{\tau=1}^{m-1} P_e ( r_\tau )
\li [ 1 - \li ( \f { r_\tau } { r_{\tau+1} } \ri )^d \ri ]\\
& = & P_e (r_i) \li [ 1 - \li ( \f{r_i}{\wh{r}_{i+1} } \ri )^d \ri ]
+ P_e ( \wh{r}_{i+1} ) \li [1 - \li ( \f{\wh{r}_{i+1} }{r_{i+1}} \ri
)^d \ri ] - P_e (r_i) \li [ 1 - \li ( \f{r_i}{r_{i+1}} \ri )^d \ri
]. \eee By virtue of the fact that $P_e ( \wh{r}_{i+1} ) \geq P_e
(r_i)$, we have \bee &  & \aleph
(\mathcal{G}_2) - \aleph (\mathcal{G}_1)\\
& \geq & P_e (r_i) \li [ 1 - \li ( \f{r_i}{\wh{r}_{i+1} } \ri )^d
\ri ] + P_e ( r_i ) \li [1 - \li ( \f{\wh{r}_{i+1} }{r_{i+1}} \ri
)^d \ri ] - P_e (r_i) \li [ 1 - \li ( \f{r_i}{r_{i+1}}
\ri )^d \ri ]\\
& = & P_e (r_i) \li [ 1 - \li ( \f{r_i}{\wh{r}_{i+1} } \ri )^d - \li
( \f{\wh{r}_{i+1} }{r_{i+1}} \ri )^d + \li ( \f{r_i}{r_{i+1}} \ri
)^d \ri ]. \eee Recall that $r_i < \wh{r}_{i+1} < r_{i+1}$, we have
\bee &  & 1 - \li ( \f{r_i}{\wh{r}_{i+1} } \ri )^d - \li (
\f{\wh{r}_{i+1} }{r_{i+1}} \ri )^d +
\li ( \f{r_i}{r_{i+1}} \ri )^d\\
& = & \f{ \wh{r}_{i+1}^d - r_i^d } { \wh{r}_{i+1}^d } - \f{
\wh{r}_{i+1}^d
- r_i^d } { r_{i+1}^d }\\
& = & \f{ ( \wh{r}_{i+1}^d - r_i^d) (r_{i+1}^d - \wh{r}_{i+1}^d) } {
\wh{r}_{i+1}^d r_{i+1}^d }\\
& > & 0. \eee It follows that $\aleph (\mathcal{G}_2) - \aleph
(\mathcal{G}_1) \geq 0$.

\epf

We are now in the position to prove statement (II) of the theorem.
For any set of grid points, we can reduce the maximal gap between
grid points by adding grid points. Every new grid point is placed at
the middle of one of the previous intervals which possess the {\it
largest} width in order to ensure that, as more grid points added,
the maximal gap of grid points tends to zero. In this process, we
create a series of nested sets of grid points $\mathcal{G}_k, \; k =
1, 2, \cd, \iy$ such that $\mathcal{G}_1 \subset \mathcal{G}_2
\subset \mathcal{G}_3 \subset \cd$. Note that \begin{eqnarray} &  &
\li | \sum_{i=1}^{m-1} P_e (r_i) \li [ 1 - \li ( \f { r_i } {
r_{i+1} } \ri )^d \ri ] -
d \int_{\f{a}{\lm}}^{a} \f{P_e(x)} {x} d x \ri | \nonumber\\
& \leq & \li | \sum_{i=1}^{m-1} P_e (r_i) \li [ 1 - \li ( \f { r_i }
{ r_{i+1} } \ri )^d \ri ] - d \sum_{i=1}^{m-1} \f { P_e (r_{i+1})
(r_{i+1} - r_i) } { r_{i+1} } \ri
| \nonumber \\
&  & + \li | d \sum_{i=1}^{m-1} \f { P_e (r_{i+1}) (r_{i+1} - r_i) }
{ r_{i+1} } - d \int_{\f{a}{\lm}}^{a} \f{P_e(x)} {x} d x \ri |
\nonumber
\\
& \leq & \f{ d (d-1) \lm (\lm -1) \varpi } { 2 a } + \f{ \lm d } {
a}
\; \sum_{i=1}^{m-1} P_e(r_{i+1})(r_{i+1} - r_i ) \la{ieq88}\\
&  & - \f{ \lm d } { a} \; \sum_{i=1}^{m-1} P_e(r_i)(r_{i+1} - r_i)
\nonumber \\
&  & + d \li | \sum_{i=1}^{m-1} \f { P_e (r_{i+1}) (r_{i+1} - r_i) }
{ r_{i+1} } - \int_{\f{a}{\lm}}^{a} \f{P_e(x)} {x} d x \ri |
\nonumber
\end{eqnarray}
where inequality (\ref{ieq88}) follows from Lemma \ref{lem99}. By
Lemma \ref{boundness}, $\bb{P}(.)$ is a continuous function with
respect to $r$. Consequently, $\f{P_e(x)} {x}$ is Riemann integrable
over interval $\li [ \f{a}{\lm}, a \ri ]$ and
\[
\lim _{\varpi \to 0} \; \sum_{i=1}^{m-1} \f { P_e(r_{i+1}) (r_{i+1}
- r_i) } { r_{i+1} } = \int_{\f{a}{\lm}}^{a} \f{P_e(x)} {x} d x.
\]
Moreover, since $P_e(x)$ is Riemann integrable, we have
\[
\lim _{\varpi \to 0} \; \sum_{i=1}^{m-1} P_e(r_{i+1})(r_{i+1} - r_i
) = \lim _{\varpi \to 0} \; \sum_{i=1}^{m-1} P_e(r_i)(r_{i+1} - r_i
) = \int_{\f{a}{\lm}}^{a} P_e(x) d x.
\]
Hence, the right hand side of inequality (\ref{ieq88}) can be made
arbitrarily small by successively cutting the gap between grid
points in half with new grid points. This proves that \[ \lim_{k \to
\iy} \aleph ( \mathcal{G}_k ) = d \int_{\f{a}{\lm}}^{a} \f{P_e(x)}
{x} d x.
\]
On the other hand, by Lemma \ref{Uencp}, we have $\aleph (
\mathcal{G}_1 ) \leq \aleph ( \mathcal{G}_2 ) \leq \aleph (
\mathcal{G}_3 ) \leq \cd$. Combining the convergency and the
monotone property of sequence $\li \{ \aleph ( \mathcal{G}_k ) \ri
\}_{k=1}^\iy$, we can conclude that $\aleph ( \mathcal{G} ) \leq d
\int_{\f{a}{\lm}}^{a} \f{P_e(x)} {x} d x$ for any set of grid points
$\mathcal{G}$. By Lemma \ref{lem889}, the expected number of rows of
the matrix of violations $V^{\bs{n}}$ is no greater than \bee 1 + N
P_e(a) + 2 N \aleph ( \mathcal{G} ) & \leq & 1 + N P_e(a) + 2 N d
\int_{\f{a}{\lm}}^{a}
\f{P_e(x)} {x} d x\\
& = & 1 + N P_e(a) + 2 N d \int_{\f{a}{\hbar}}^{a} \f{P_e(x)} {x} d
x \eee for any $\mathcal{G}$. Such bound applies to any $V^j$
because the number of rows of $V^j$ is non-decreasing with respect
to $j$. Finally, the inequality of (\ref{ieq_encp}) can be proved by
making use of the observation that $P_e(x) \leq P_e(a), \; \fa x \in
\li [ \f{a}{\hbar }, \; a \ri ]$.

\eed

\section{Proof of Theorem \ref{CI_band}}

We need the following lemma, which has recently been obtained in \cite{C3}.

\begin{lemma} \label{Tape_Massart} Let $X_i, \; i = 1, \cd, N$ be i.i.d.
Bernoulli random variables such that $\Pr \{ X_i = 1 \} = 1 - \Pr \{ X_i = 0 \} = \bb{P}_X > 0$. Let $K =
\f{\sum_{i = 1}^N X_i} {N}$. Then
\[
\Pr \left\{ \mathcal{L} (K)< \bb{P}_X < \mathcal{U} (K) \right\}
>1 - \delta.
\]
\end{lemma}

Applying Lemma \ref{Tape_Massart}, we have $\Pr \{ \mathcal{L}
(K_{i+1}) < \bb{P} (r_{i+1}) < \mathcal{U} (K_{i+1}) \} > 1 -
\f{\de}{2}$ and $\Pr \{ \mathcal{L} (K_i) < \bb{P} (r_i) <
\mathcal{U} (K_i) \}
>1 - \f{\de}{2}$. Hence by the Bonferroni's
inequality,
\[
\Pr \{ \mathcal{L} (K_{i+1}) < \bb{P} (r_{i+1}) < \mathcal{U}
(K_{i+1}), \qu \mathcal{L} (K_i) < \bb{P} (r_i) < \mathcal{U}
(K_i)\} > 1 - \de.
\]
By the definitions of $\bb{P}^*(r), \; \overline{\bb{P}}(r)$ and $
\underline{\bb{P}}(r)$, we have that event $\{ \mathcal{L} (K_{i+1})
< \bb{P} (r_{i+1}) < \mathcal{U} (K_{i+1}), \; \mathcal{L} (K_i) <
\bb{P} (r_i) < \mathcal{U} (K_i)\}$ implies event $\{
\underline{\bb{P}}(r) + \varsigma < \bb{P}^*(r) <
\overline{\bb{P}}(r) - \varsigma, \; \fa r \in [r_i, r_{i+1}] \}$.
Hence, $\Pr \{ \underline{\bb{P}}(r) + \varsigma < \bb{P}^*(r) <
\overline{\bb{P}}(r) - \varsigma, \; \fa r \in [r_i, r_{i+1}] \}
>1 - \de$. By Theorem \ref{bound1} and the gridding scheme, $\Pr \{ |
\bb{P}^*(r) - \bb{P}(r) | < \varsigma, \; \fa r \in [r_i, r_{i+1}]
\} = 1$. Applying Bonferroni's inequality, we have \be \la{am} \Pr
\{ \underline{\bb{P}}(r) + \varsigma < \bb{P}^*(r) <
\overline{\bb{P}}(r) - \varsigma, \; | \bb{P}^*(r) - \bb{P}(r) | <
\varsigma, \; \fa r \in [r_i, r_{i+1}] \}
>1 - \de.
\ee Finally, the theorem is proved by observing that the left hand side of
inequality (\ref{am}) is no greater
than $\Pr \{ \underline{\bb{P}}(r) < \bb{P}(r) < \overline{\bb{P}}(r), \;
\fa r \in [r_i, r_{i+1}] \}$.


\begin{thebibliography}{10}

\bibitem{bai}E. W. BAI, R. TEMPO, AND M. FU,
``Worst-case properties of the uniform distribution and randomized algorithms for robustness analysis,'' {\it
Mathematics of Control, Signals and Systems}, vol. 11, pp.183-196, 1998.

\bibitem{BLT} B. R. BARMISH, C. M. LAGOA, AND R. TEMPO,
``Radially truncated uniform distributions for probabilistic
robustness of control systems,'' {\it Proc. of American Control
Conference}, pp. 853-857, Albuquerque, New Mexico, June 1997.


\bibitem{BL}
B. R. BARMISH AND C. M. LAGOA, ``The uniform distribution: a
rigorous justification for its use in robustness analysis,'' {\it
Mathematics of Control, Signals and Systems}, vol. 10, pp.
203-222, 1997.

\bibitem{Barmish2} B. R. BARMISH AND P. S. SHCHERBAKOV, ``On avoiding
vertexization of robustness problems: The approximate feasibility concept,'' {\it IEEE Transactions on Automatic
Control}, vol. 42, pp. 819-824, 2002.


\bib{CJZ} X. CHEN, J. ARAVENA, AND K. ZHOU, ``Risk analysis in robust
control --- making the case for
probabilistic robust control,'' {\it Proc. of American Control
Conference}, pp. 1533-1538, Portland, Oregon, June 2005.

\bibitem{C0} X. CHEN, K. ZHOU, AND J. ARAVENA, ``Fast construction of
robustness degradation function,'' {\it SIAM Journal on Control
and Optimization}, vol. 42, pp. 1960-1971, 2004.


\bibitem{C3} X. CHEN, K. ZHOU AND J. ARAVENA, ``Fast universal
algorithms for robustness analysis,'' {\it Proceedings IEEE Conference on Decision and Control}, pp. 1926-1931,
Maui, December 2003.

\bib{Chernoff} H. CHERNOFF, ``A measure of asymptotic
efficiency for tests of a hypothesis based on the sum of
observations,'' {\it Annals of Mathematical Statistics}, vol. 23, pp.
493-507, 1952.

\bibitem{Clo}
C. J. CLOPPER AND E. S. PEARSON, ``The use of confidence or
fiducial limits illustrated in the case of the binomial,''
{\it Biometrika}, vol. 26, pp. 404-413, 1934.

\bib{Feller} W. FELLER, {\it An Introduction to Probability Theory and Its
Applications}, Wiley, 1968.


\bib{PT} P. F. HOKAYEM AND C. T. ABDALLAH, ``Quasi-Monte Carlo methods in
robust
control design,'' {\it Proceedings IEEE Conference on Decision and
Control}, pp. 2435-2440, Maui, December 2003.

\bib{Kan} S. KANEV, B. De SCHUTTER, AND M. VERHAEGEN, ``An ellipsoid
algorithm for probabilistic robust controller design,'' {\it
Systems and Control Letters}, vol. 49, pp. 365-375, 2003.



\bib{Ko} V. KOLTCHINSKII, C.T. ABDALLAH, M. ARIOLA, P. DORATO, AND D.
PANCHENKO, ``Improved sample complexity estimates for statistical learning control of uncertain systems,'' {\it
IEEE Transactions on Automatic Control}, vol. 46, pp. 2383-2388, 2000.

\bibitem{BLT2} C. M. LAGOA, ``Probabilistic enhancement of classic
robustness margins: a class of none symmetric distributions,'' {\it
Proc. of American Control Conference}, pp. 3802-3806, Chicago,
Illinois, June 2000.


\bib{La3} C. M. LAGOA, X. LI, M. C.
MAZZARO, AND M. SZNAIER, ``Sampling random transfer functions,''
{\it Proceedings IEEE Conference on Decision and Control}, pp.
2429-2434, Maui, December 2003.


\bibitem{MS}
C. MARRISON AND R. F. STENGEL, ``Robust control system design
using random search and genetic algorithms,'' {\em IEEE
Transaction on Automatic Control}, vol. 42, pp. 835-839, 1997.


\bibitem{RS}
L. R. RAY AND R. F. STENGEL, ``A monte carlo approach to the
analysis of control systems robustness,'' {\it Automatica}, vol.
3, pp. 229-236, 1993.

\bibitem{SB}
S. R. ROSS AND B. R. BARMISH, ``Distributionally robust gain
analysis for systems containing complexity,'' {\it Proceedings of
Conference on Decision and Control}, pp. 5020-5025, Orlando,
Florida, December 2001.

\bibitem{SR}
R. F. STENGEL AND L. R. RAY, ``Stochastic robustness of linear
time-invariant systems,'' {\it IEEE Transaction on Automatic
Control}, vol. 36, pp. 82-87, 1991.


\bib{Wang} Q. WANG AND R. F. STENGEL, ``Robust control of nonlinear
systems with parametric uncertainty,'' {\it Automatica}, vol. 38,
pp. 1591-1599, 2002.


\end{thebibliography}
\end{document}